\documentclass[12pt, preprint]{aastex}

\usepackage{emulateapj5}
\usepackage{graphicx}
\usepackage{fancyheadings}
\usepackage{color}

\newcommand{\gae}{\mathrel{\raise .4ex\hbox{\rlap{$>$}\lower 1.2ex\hbox{$\sim$}
} }}

\newcommand{\pks}{PKS~1127-145}
\shorttitle{X-ray Jet in PKS~1127-145}

\slugcomment{\today}

\shortauthors{Siemiginowska et al.}

\begin{document}

\title{The 300~kpc Long X-ray Jet in PKS~1127-145, $z=1.18$ Quasar: Constraining
X-ray Emission Models.}

\author{Aneta Siemiginowska$^1$, \L ukasz Stawarz$^{2, \, 3,\, 5}$, C.C. Cheung$^{4,5}$,
D.E. Harris$^1$, Marek Sikora$^6$, Thomas L. Aldcroft$^1$, Jill Bechtold$^7$}
\affil{$^1$ Harvard-Smithsonian Center for Astrophysics, Cambridge, MA 02138\\ 
$^2$ Landessternwarte Heidelberg, K\"onigstuhl, and Max-Planck-Institut f\"ur 
Kernphysik, Saupfercheckweg 1, 69117 Heidelberg, Germany\\
$^3$ Obserwatorium Astronomiczne UJ, ul. Orla 171, 30-244 Krak\'ow, Poland\\
$^4$ Jansky Postdoctoral Fellow, National Radio Astronomy
Observatory.\\
$^5$ Kavli Institute for Particle Astrophysics and Cosmology, Stanford
University, Stanford, CA 94305\\
$^6$ Copernicus Astronomical Center, ul. Bartycka 18, 00-716 Warsaw, Poland\\
$^7$ Steward Observatory, University of Arizona, 933 North Cherry Avenue, Tucson, AZ 85721}

\email{asiemiginowska@cfa.harvard.edu}

\begin{abstract}

We present a $\sim100$~ksec {\it Chandra} X-ray observation and new
VLA radio data of the large scale, 300~kpc long X-ray jet in \pks, a
radio loud quasar at redshift z=1.18. With this deep X-ray observation
we now clearly discern the complex X-ray jet morphology and see
substructure within the knots. The X-ray and radio jet intensity
profiles are seen to be strikingly different with the radio emission
peaking strongly at the two outer knots while the X-ray emission is
strongest in the inner jet region. The jet X-ray surface
brightness gradually decreases by an order of magnitude going out from
the core. The new X-ray data contain sufficient counts to do spectral
analysis of the key jet features.  The X-ray energy index of the
inner jet is relatively flat with $\alpha_X = 0.66\pm0.15$ and steep
in the outer jet with $\alpha_X = 1.0\pm0.2$. We discuss the
constraints implied by the new data on the X-ray emission models and
conclude that ``one-zone'' models fail and at least a two component
model is needed to explain the jet's broad-band emission. We propose
that the X-ray emission originates in the jet proper while the bulk of
the radio emission comes from a surrounding jet sheath.  We also
consider intermittent jet activity as a possible cause of the observed
jet morphology.

\end{abstract}


\section{Introduction} 

Jets span distances of hundreds of kpc to Mpc and constitute the largest
physical manifestation of the AGN phenomenon.  However, fundamental
questions about the nature of jets remain unanswered, while emission
processes associated with the production of X-rays and $\gamma$-rays
are critical to understanding quasars.  The discovery of many X-ray
jets over the last 6 years with the {\it Chandra} X-ray Observatory
\citep{wei02} ({\it Chandra} hereafter) indicate that jets, and in
particular large scale jets ($>100$~kpc) are common (e.g. Schwartz et
al 2000, Marshall et al 2001, Worrall et al 2001, Sambruna et al 2002,
Siemiginowska et al 2003, Sambruna et al 2004, Marshall et al 2005)

The synchrotron origin of the radio to optical jet emission has now
been well established.  However the origin of the jet X-ray emission
is puzzling (see Harris \& Krawczynski 2006 for review) since a
straight extrapolation of the synchrotron (radio-to-optical) continuum
into the X-rays severely underpredicts the luminosity of powerful {\it
Chandra} large scale jets. Thus, the same single power-law population
of electrons cannot produce the radio, optical, and X-ray emission in
the framework of a homogeneous one-emission zone approximation (see
for example Sambruna et al 2004).  The synchrotron self-Compton (SSC)
process cannot easily explain the data because it does not produce
enough X-rays at the equipartition fields and therefore requires large
departures from the minimum-power condition (Chartas et al 2000,
Harris \& Krawczynski 2002, Kataoka \& Stawarz 2005). It was proposed
that the X-rays from large-scale quasar jets might be associated with
inverse-Compton scattering of the Cosmic Microwave Background (IC/CMB)
photons implying large jet bulk Lorentz factors ($\Gamma$) at hundreds
of kpc from the active nuclei (Tavecchio et al. 2000, Celotti et
al. 2001, Schwartz 2002). However, recent {\it Chandra} observations
indicate several problems with this simple model: (1) the observed
offsets between X-ray and radio peak brightnesses in some jet knots,
with the X-ray peak being closer to the core; (2) the large variety of
X-ray spectral indices observed in several jets; (3) the spectral
dependence of the knot shapes and profiles, as well as possible knot
substructure (see Harris \& Krawczynski 2006 for a wide discussion).

The most widely discussed types of models fall into two main classes:
(i) synchrotron scenarios invoking separate populations of radiating
electrons and/or non-standard broad-band electron spectra (Dermer \&
Atoyan 2004, Stawarz et al 2004) and (ii) more complex versions of the
inverse-Compton/CMB models involving knot inhomogeneity or jet
deceleration (Tavecchio et al 2003, Georganopoulous \& Kazanas
2004). The most recent {\it Spitzer} and {\it Chandra} observations of
the famous 3C~273 jet reported by Uchiyama et al. (2006) and Jester et
al. (2006) seem to favor the former (i.e. synchrotron) scenario,
revealing peculiar spectral shapes of the
\emph{synchrotron} (polarized) knot continua in the IR-to-UV photon
frequency range, and characteristic broad-band spectral changes along
the outflow. Both of these are consistent with the `two-electron
population'/`non-standard particle spectra' synchrotron model, but
could not be explained in the framework of the inverse-Compton
scenario. Yet the question remains if the 3C~273 jet is representative
of other large-scale quasar jets detected by {\it Chandra}, or is a
unique source.

A possible discriminant between synchrotron and IC/CMB emission models
in all the sources is related to significant differences in lifetimes
of the radiating particles, which will manifest itself in the
observed jet morphologies at different wavelengths. In particular,
X-rays emitting synchrotron electrons (with Lorentz factors $\gamma \geq
10^7$) have much shorter lifetimes than the lower energy radio
emitting electrons and hence one may expect the X-ray emission from
knots to be more compact than the corresponding radio
emission. Instead the electron cooling times related to the IC/CMB
emission are long since the X-rays are produced by a low energy
population of electrons ($\gamma <300$) thus implying continuous,
intra-knot X-ray emission outside the bright knots. Such an idealized
picture may be however significantly complicated by
intermittent/modulated jet activity and efficient particle
acceleration acting within the entire jet volume (Stawarz et
al. 2004). In such a case, the multiwavelength morphology of the
outflow is controlled by the jet activity and acceleration timescales
rather than by the radiative cooling timescales of the particles.

The differences in the lifetimes of the radiating electrons could also
shape the spectral profiles along the jets, i.e. runs of both
broad-band (e.g., radio-to-optical) power-law slopes, and also of the
narrow-band (e.g., radio, or X-ray) spectral indices. However, in
order to extract such informations from the data, deep multiwavelength
observations are needed. We note that Sambruna et al (2004) found very
flat ($\alpha_{X} \sim 0.5$) to almost inverted ($\alpha_{X} \sim
0.1$) spectra in a few X-ray jets (though with large uncertainties -
the majority have errors from 0.3-0.9), which they associated with the
IC/CMB from the electron population near the low energy cut-off of the
electron distribution. Alternatively, in the framework of the
synchrotron scenario, flat X-ray continua may indicate spectral
pile-up occurring at the high-energy part of the electron energy
distribution, which is in fact expected if the continuous particle
acceleration acts efficiently enough (see Stawarz \& Ostrowski 2002).

One of the longest X-ray jets known\footnote{see 
\texttt{http://hea-www.harvard.edu/XJET/}} 
was discovered by {\it Chandra} in the second year of the mission
(Siemiginowska et al. 2002, Paper I). With a length of almost 300~kpc
projected onto the sky, this jet, associated with the redshift z=1.18
radio-loud quasar \pks\,, poses several questions for X-ray emission
models.  The one-sided jet shows an X-ray surface brightness which
declines with the distance from the core, while the radio brightness
increases. The jet slightly curves, with position angle (with respect
to the core) changing from $\sim 64 \deg$ at the core to $\sim 43
\deg$ at the jet's end.  The discovery observation identified three
main knots along the jet with the furthest knot C being the
weakest. The prominent knots A and B are connected by continuous X-ray
emission that stops beyond knot B at $\sim 22$~arcsec from the core.
The VLA radio maps show very low brightness emission along the jet
(Fig.5 in Siemiginowska et al 2002) and the X-ray to radio intensity
ratio decreases along the jet. The HST WFPC2 observation gives upper
limits to the knots optical brightness, but they are too high for
constraining the emission models.  Thus only radio and X-ray
observations can be used to study emission processes in this jet.

The new {\it Chandra} data significantly improve on the discovery
observation.  With the deep exposure and high S/N it is obvious that
the jet morphology is complex and some substructure is clearly visible
in the large knots. We now do not detect offsets of 0.8-2~arcsec
between the X-ray and radio peak brightnesses reported in Paper I, but
conclude that they are now mostly attributable to the jet
substructure. We also obtain well constrained spectral information for
the main jet features.  We discuss the results in the context of
currently considered emission models and pose questions. This is
clearly a remarkable jet providing us with many constraints on the
current emission models. Below we discuss in detail ``one-component''
emission models and show that they fail to explain the \pks\ jet
data. We suggest that a two-component model (the proper jet and a
sheath) can describe the data. We also consider the possibility of
modulated jet activity, which might be responsible for the observed
jet morphology.

The X-ray data are presented in Section~\ref{section-chandra}, radio
data in Section~\ref{section-radio}, jet morphology and spectral
properties are presented in Sections~\ref{section-morphology}
and~\ref{section-spectra}, optical data in Section 6, and a discussion
of the results and implications for jet models is in
Section~\ref{section-models}.  Throughout this paper we use the
cosmological parameters based on the WMAP measurements (Spergel et
al. 2003): H$_0=$71~km~sec$^{-1}$~Mpc$^{-1}$, $\Omega_M = 0.27$, and
$\Omega_{\rm vac} = 0.73$. At $z= 1.18$, 1\arcsec\ corresponds to
$\sim$8.3~kpc.  For a power law radiation spectrum we adopt the
convention: flux density, $S_\nu \propto \nu^{-\alpha}$.

\section{Chandra Observations}
\label{section-chandra}

The new observation of \pks\, with {\it Chandra} was obtained on April
25-27, 2005 (obsid 5708) with $\sim$106~ksec exposure. This was the
second {\it Chandra} observation of this quasar following the
discovery of the X-ray jet in AO1 pointing (30~ksec, obsid=866). The
source was placed 30~arcsec from the default aim-point position on the
ACIS-S backside illuminated CCD, chip S3 (Proposers' Observatory
Guide,
POG\footnote{http://cxc.harvard.edu/proposer/POG/index.html}). To
reduce the effects of pileup we collected the data in 1/8 subarray
readout mode of one CCD only, which resulted in a 0.441 sec frame
readout time.  The quasar's count rate of 0.62~cts~s$^{-1}$ gives
8-10$\%$ pileup in the core for this choice of the readout mode (see
POG). After standard filtering, the effective exposure time for this
observation was 103,203 sec.

The standard off-set pointing was used in this observation to avoid
the node boundary. However, due to the drift of the optical axis since
the {\it Chandra} launch in 1999 the aim point location has moved
closer to the node boundary (POG) and the quasar core was crossing the
node during the observation. This affected the core data and about
30\% of the core photons were marked as bad due to their location on
the node boundary. Because of a choice of the roll angle the jet was
placed away from the node and the jet data more than 3~arcsec away
from the quasar centroid are not affected by the node boundary.

The X-ray data analysis was performed with the CIAO~3.3
software\footnote{http://cxc.harvard.edu/ciao/} using calibration
files from the CALDB~3 database.  We ran {\tt acis\_process\_events}
to remove pixel randomization and to obtain the highest resolution
image data. The X-ray position of the quasar (RA$=11^h30^m07^s.05$, 
DEC$=-14^d49^m27^s.3$ J2000) agrees with the radio position~\citep{john1995} 
to better than 0.25$\arcsec$,
(which is smaller than {\em{Chandra}}'s 90$\%$ pointing accuracy of
0.6\arcsec, Weisskopf et al. 2003). All spectral modeling was done in
{\it Sherpa} \citep{fre01}.

\section{VLA Observations}
\label{section-radio}

The VLA observations of Feb 2001 from program AH730 (1.4 and 8.5 GHz) are
described in Paper I, but are also included in Table~\ref{tab-radio} for
completeness.  The 1.4 GHz data were rummaged here with a smaller (uniform
weighted) beam revealing the presence of the X-ray detected feature O. This
is the only radio detection of this knot at the sensitivity of the higher
frequency VLA maps (below).

We also reduced 5 GHz observations obtained as part of this program (June
2001) which were not available at the time of submission of Paper I. These
data were taken in a less extended (B) configuration to synthesize matched
resolution images at different frequencies (Figure~6). All data were
calibrated in AIPS\citep{bri94} and imaging was done in DIFMAP \citep{she94}.
Four antennas were sub-arrayed to another project as our observations
commenced, so effectively, only 23 antennas participated in this experiment.

New longer integration 8.5 GHz data were obtained in May 2005 mainly
to improve the sensitivity in order to better define the radio jet
morphology than in the previous 2001 image published in Paper I. Three
antennas were being retrofitted for the EVLA at the time leaving 24 in
the new experiment. The first hours of observations were flagged as
this was when the target was rising (15-25 deg elevation) and high
cloud cover was reported. This left 6.25 hours of on-source time in
the 8 hr observing run.  After standard calibration and imaging, we
removed the core (3.23 Jy) with the UVSUB task in AIPS to isolate the
inner jet emission. We merged the new 8.5 GHz dataset with the
previous one from 2001 but found no significant improvement in the
images produced; only data from the latter epoch are used here. In the
final core-subtracted image clean components were convolved with a
circular Gaussian of FWHM=0.7'', so radio structures appearing
non-circular in Figures~7 \&~8 are intrinsically elongated.

Off-source rms levels for the radio images are listed in Table~1 along with
the resultant dynamic range relative to the image peaks. In all cases, the
rms is higher than the expected theoretical limit (factors from $\sim$3--10
times larger). This is presumably due to a combination of the bright radio
core and the low-declination of the target making the field difficult to
map. In the vicinity of the core, imaging artifacts (in the north-south
direction) are somewhat obvious and the rms levels are even higher than the
off-source values. This is reflected in the upper limits for several of the
undetected inner radio knots reported in Table~2.

\section{Jet Morphology}
\label{section-morphology}

\subsection{X-ray Morphology}

We consider the X-ray morphology of the large scale jet starting
$\sim$~3~arcsec away from the quasar centroid. The analysis of the
quasar core and the small scale structure in the vicinity of the quasar
will be presented elsewhere.
 
The smoothed ACIS-S image in Fig.~\ref{smooth} shows a complex
structure of a continuous X-ray emission along the jet up to $\sim
22$~arcsec distance from the core.  There are several brightenings
present within a diffuse X-ray emission between the brightest
knots. The jet width is resolved and a structure at $\sim 8$~arcsec
distance from the core showing a kink is ``real''. We discuss this
kink in the section below.

We constructed an X-ray jet profile along the jet and a control
profile with the regions shown in Fig.~\ref{jet-reg}. The background
for each jet's cell was taken from an annulus with radial widths
of the jet cell, excluding the readout streaks, the control region and
a bright point source. The resulting profile is presented in
Fig.~\ref{profile-hr}. The jet X-ray emission declines gradually along
the jet with the knots A, B and C present. Figure~\ref{profile-zoom}
is a zoomed version of Fig.~\ref{profile-hr} to present the emission
beyond knot B. The X-ray jet emission seems to stop at the ``edge'' of
knot B before it slightly increases to form knot C.  Comparison
between the jet and control profiles in Fig.~\ref{profile-zoom}
indicates that the emission of knot C ends at $\sim 30$~arcsec,
although the emission within the final 4~arcsec (between 26-30'') is only
marginally detected.

The morphology of each knot is quite different. The profile of knot A
shows a relatively sharp rise and a plateau before a gradual decline,
while knot B profile is more peaked with a strong rise and decline
(see Fig.~\ref{profile-zoom}). Knot C is the faintest knot with the
X-ray emission spread over a larger area. As we show below, X-ray
emission covers the area of the strongest radio emission, which is
also diffuse here. One gets an impression that knot C looks more like
a lobe than a knot. Although, even with {\it Chandra} resolution we
still can only resolve structures on $\sim 1$~arcsec scales and even
``compact'' knots A and B are of the size of the entire M87 jet for
example (see Fig.6 in Harris \& Krawczynski 2006 for comparison of the
jet projected lengths). Thus what we see reflects the overall
integrated intensity within each region, while there might be many
smaller scale regions which cannot be resolved.

Looking at both the jet profile and the smoothed X-ray image it is
clear that the morphology of the jet is complex. In addition to the
three knots we identify brightenings (knots O and I) between the core
and knot A, as well as a drop in the surface brightness at $\sim
8.5$~arcsec from the core.  We also detect the emission in between the
knots.

\subsection{Radio Morphology}

The VLA radio maps 1.4~GHz and 5~GHz are presented in
Fig.~\ref{radioCL}. The VLA 8.5~GHz map is in Fig.~\ref{radio-X1} and
Fig.~\ref{radio-X2} and in this map the core has been subtracted to
highlight the very faint jet structure.  It is obvious that the radio
emission of the innermost jet is very faint, with knot A at the
detection threshold. The emission becomes strong at the outermost
regions with knots B and C being quite strong.  There is continuous
emission between the last two knots, so their peak brightnesses are
connected.

The 8.5~GHz radio profile of the jet shown in Fig.~\ref{radio-X-profile}
shows the jet brightness increasing towards the final knot. The
brightness of the last two knots is almost identical.  The innermost
regions of the jet, identified as region I are also detected at this
frequency while there is only an upper limit to the radio emission at
the location of the O feature.

Table~\ref{tab1} presents the radio flux density measurements for each
map.  Given the three radio frequencies we can determine the radio
spectral index for the brightest knots and get $\alpha_r$ equal to 0.8
and 0.9 for knots B and C respectively.

\subsection{Radio and X-rays Connection.}

Figure~\ref{radio-X-profile} compares X-ray and radio intensities
along the jet.  The profile intensities were extracted assuming the
same box region for the radio 8.5~GHz and X-ray (0.3-7~keV) data
(using {\tt ds9 projection}) and taking 2.5~arcsec width for the
region box. The two profiles look quite different.  The X-ray
intensity decreases with the knots on top of a continuous emission,
while the radio intensity stays at a very low level to about 16~arcsec
when it sharply rises to form knot B, than decays to a diffuse level
and rises again to form knot C.  The three knots I, A and B are the
strongest X-ray features, while knots B and C dominate the radio
emission.

The X-ray emission of knot A is very prominent in comparison to the
radio emission. It does not have an obvious peak, but rather a broad
flat profile. Knot A is very faint in radio and its radio spectrum is
very steep ($\alpha_r > 1.35$) in comparison to knots B and C whose
spectra are more typical, $\sim 0.8-0.9$ (see Table~\ref{tab1}).

The emission from knot B is very prominent in both radio and X-rays
and the knot profiles can be studied in more details.  In Paper I we
reported the offsets between the radio and X-ray peak brightnesses of
1.4~arcsec for this knot (using the radio data with the uncertainties
of 1~arcsec at 1.4~GHz and X-rays with uncertainties of 0.3~arcsec).
Now, with the improved S/N {\it Chandra} image and 8.5~GHz radio data
we can model directly the profiles of the knot and obtain more
accurate locations of the peaks (see Fig.~\ref{radio-X-profile}). We
measure that the radio (8.5~GHz) and X-ray peak brightnesses are
coincident to within 0.2~arcsec.  The X-ray emission precedes the
radio emission for this knot in the sense that the X-rays start rising
before the radio and decay earlier than the radio.  Thus the X-ray
profile of this knot seems to be shifted in comparison to the radio
profile.

The radio emission of knot C is strong and its profile shows a step
before the final rise to the maximum at 28~arcsec. Then the emission
drops and the knot ends at 31~arcsec. The X-ray emission is very faint
with constant intensity.

\section{Spectral Properties of the X-ray Knots}
\label{section-spectra}

Fig.~\ref{profile-hr} show the hardness ratio along the jet. We define
the hardness ratio as (H-S/H+S), where H is the hard source counts
with 2-7~keV energies, and S is the soft source counts with 0.5-2~keV
energies. The hardness ratio errors are calculated using the Bayesian
algorithm of Park et al (2006). The jet's X-ray spectrum has a
constant hardness ratio up to about 8~arcsec, in the vicinity of the
``kink'' with a sudden drop in the jet intensity. The hardness ratio
suggests a spectral change at this point which continues further along
within knot A.  It seems that the jet spectrum hardens along knot A,
and then becomes softer again before hardening within knot B. The
Signal-to-Noise ratio is too low to study the hardness ratio
variations beyond knot B.

To characterize the jet spectral properties we extracted the X-ray
spectra of small parts of the jet, assuming the box regions shown
in Fig.\ref{jet-reg}.  The total number of counts and the best fit power
law model parameters are presented in the Table~\ref{tab1}.  The power
law slope indicates a hard (flat) spectrum $\alpha_{X}
\sim 0.6-0.7$ for the inner regions I, O and knot A and a soft
(steep) spectrum, $\alpha_{X} \sim 1-1.2$, for the outer knots B
and C. 

To check whether the difference in the X-ray emission from the
inner and outer jet is significant we extracted the spectra from the
two large regions: the inner jet region contains knots I,O and A, and
the outer jet knots B and C. We then fit these two spectra (using the
calibration files appropriate for the new extraction regions) with an
absorbed power law model. We obtain the power law slopes for each
region, $\alpha_{X,inner} = 0.64\pm0.07$ and $\alpha_{X,outer} =1.29
\pm0.22 $. Thus the spectral difference between the two regions 
is at $\sim 2.8\sigma$ level. The hardness ratios based on the source
counts for the inner and outer jets are -0.48$^{0.03}_{-0.04}$ and
$-0.74^{0.10}_{-0.08}$ respectively.

\subsection{X-ray Diffuse Emission}

We analyzed the X-ray emission between the knot A and B regions,
e.g. in the annulus with inner and outer radii of 14 and 16.5
arcsec. There are total of 24 counts with energy between 0.3-8 keV and
18.3$\pm5.4$ net counts (assuming a background annulus with the same
radii, but excluding the jet region). The detected number of counts is
too small for detailed spectral analysis. However, the spectrum is
soft based on the hardness ratio of -0.73$^{+0.21}_{-0.15}$.

\section{Optical Jet?}

An optical image of \pks\ that was taken with HST \citep{sie02} shows a
number of galaxies with a large galaxy across the X-ray jet. A Damped
Lyman-$\alpha$ system (DLA) at redshift 0.312 has been described by
\cite{bergeron1991} and \cite{bechtold2001} (see also
\cite{bechtold2002} for detailed optical spectra). Because the DLA
galaxy has to be located close to the line of sight to the quasar, it
may affect the emission from the quasar jet. Our comparison between
the X-ray and radio jet morphology indicates that the change in the
jet shape is present in both bands (see Fig.~\ref{smooth} and
Fig.~\ref{radioCL}), thus we can exclude the absorption of jet
emission by the foreground galaxy as a cause for the observed jet
morphology. A slight change in the direction of the jet might
therefore be intrinsic to the jet or a result of gravitational lensing
of the jet by the foreground galaxy.

There is no detection of the optical jet and the limits for the
optical jet emission were given in
\citet{sie02} for knots B and C. The limits for knot A and remaining inner knots 
cannot be obtained due to the galaxies in the field. These limits are
too high to provide additional constraints on the jet
properties. Deeper optical/IR imaging is necessary to improve the
limits.

\section{Discussion}
\label{section-models}

The radio and X-ray images of the PKS 1127-145 jet and the analysis of the 
jet's broad-band spectral properties reveal significant differences between
the inner ($< 14''$) (knots I, O, and A; hereafter `the inner jet'),
and the outer portions of the jet ($16''-30''$, knots B and C;
hereafter `the outer jet').  This is not unusual among the large-scale
quasar jets and it is seen in the 3C~273 jet for example (Uchiyama et al. 2006
and references therein). However, one should note the large
extension of the PKS 1127-145 jet: both its inner and outer parts have
projected sizes of $\sim 100$ kpc each, while the entire 3C 273 jet
structure is only $\sim 60$ kpc-long.

The PKS~1127-145 inner jet is very weak in the radio band, especially at
high frequencies ($8.5$~GHz), while it is bright in X-rays. However, 
the X-ray and radio luminosities of the outer jet are roughly comparable
(see figures 5, 7, and 9). A general trend of a smooth decrease in the 
X-ray flux, and a slow increase in the radio flux can be noted along the 
jet, with the exception of the innermost knot I (for which the measured 
radio fluxes may suffer from contamination of the extremely bright core). 
This, discussed previously by Hardcastle (2006), is again reminiscent 
of the 3C 273 jet.  In particular, the ratio of $1$ keV to
$1.4$ GHz luminosities decreases by about two orders of magnitude starting 
from $[\nu L_{\nu}]_{\rm 1 \, keV} / [\nu L_{\nu}]_{\rm 1.4 \, GHz} 
\approx 50$ at the position of knot O, to $\sim 0.5$ at the
position of knot C. At high radio frequencies this decrease
seems to be even stronger, since the upper limits for the $8.5$~GHz
flux for knots O and A translates to the luminosity ratio 
$[\nu L_{\nu}]_{\rm 1 \, keV} / [\nu L_{\nu}]_{\rm 8.5 \, GHz} > 100$.

The X-ray spectral index is constant within the inner jet, $\alpha_{X}
\approx 0.65-0.7$, and it is slightly smaller than the radio spectral indices
between $1.4$, $5$ and $8.5$ GHz, namely $\alpha^{1.4}_{5}, \,
\alpha^{5}_{8.5} > 1.3$. Also it is smaller than the radio-to-X-ray
power-law slopes (see figure 10), although the errors in estimating
the appropriate radio fluxes are large. We note that at the position
of knot A the $8.5$ GHz flux is especially low and the high frequency
part of the radio continuum is very steep. Thus, the electrons
emitting X-ray photons in the inner jet, regardless of the particular
emission process (synchrotron or inverse-Compton), are characterized
by a much flatter energy spectrum than the radio-emitting ones.  The
situation is however much different for the outer jet. Here the
relatively large X-ray spectral indices, $\alpha_{X} \approx 1-1.2$,
seem to be larger than the radio-to-X-ray power-law slopes or radio
spectral indices, which are all $\lesssim 1$ (see figure 10).

Lack of reliable optical data makes it difficult to investigate the
broad-band energy distribution of the inner jet emission. On the other
hand, as shown recently by Uchiyama et al. (2006) and Jester et al. (2006)
for the case of the 3C 273 jet, a comparison between the broad-band
radio-to-optical and the optical-to-X-ray power-law slopes \emph{can be 
misleading} in deducing the origin of the jet X-ray emission 
\emph{unless detailed optical photometry}, and not solely
upper limits, is available. Such detailed optical information is
unfortunately not available for the PKS 1127-145 jet, and so below we
focus strictly on the confrontation of the radio and X-ray data, in
particular their
\emph{global} characteristics, instead of modeling
isolated knot regions.  First we discuss whether the observed
morphological and spectral properties of the radio/X-ray PKS 1127-145
jet can be explained in the framework of the `one emitting-zone'
IC/CMB and synchrotron models. The particular issues to be explained
are: (i) a decrease of the X-ray flux, (ii) an increase of the radio
flux, (iii) a steepening of the X-ray continuum, and (iv) a flattening
of the radio continuum along the jet.  Later, in the last two
sub-sections, we present a two-component jet emission model and
finally discuss the application of intermittent/modulated jet activity
to this source.

\subsection{The `One-Zone' Inverse-Compton Hypothesis}

The simplest version of the one-zone IC/CMB model to consider would 
involve relativistic outflow with negligible bulk deceleration and 
energy dissipation, expanding adiabatically when propagating through
the intergalactic medium. This, in fact, was argued by Tavecchio et al. 
(2004) to be the case for the {\it Chandra} jets in PKS 1510-089 and
1641+399, and so below we check if it also applies to PKS 1127-145 source. 
Under the above assumptions, the appropriate inverse-Compton luminosity 
of the jet plasma with the comoving volume $V'$ observed at a given frequency 
$\nu_{ic} \sim \delta^2 \, \nu_{cmb} \, \gamma^2$ can be written as
\begin{equation}
[\nu L_{\nu}]_{ic} \sim \delta^6 \, \Gamma^{-2} \, c \, 
\sigma_T \, V' \, U'_{cmb} \, \left[\gamma^3 \, n'(\gamma)\right]_{\gamma 
= \sqrt{\nu_{ic} / \nu_{cmb} \, \delta^2}} \quad ,
\end{equation}
\noindent
where $\Gamma$ and $\delta$ are the jet Lorentz and Doppler factors, 
$U'_{cmb} = \Gamma^2 \, U_{cmb}$ is the jet comoving energy density 
of the CMB radiation, and $n'(\gamma)$ is the electron energy 
distribution such that the electron energy density is $U'_e = \int 
\gamma \, m_e c^2 \, n'(\gamma) \, d\gamma$ (see, e.g., Stawarz et al. 
2003). Here the unprimed quantities correspond to the stationary rest 
frame at the redshift of the source. The adiabatic losses due to the 
jet's expansion between the location $r_0$ and $r$ from the nucleus
leads to $N'[\gamma, \, r] = \left(r / r_0\right)^A \, N'[\gamma_0, 
\, r_0]$ and $\gamma = \left(r / r_0\right)^{-A} \, \gamma_0$,
where $N'[\gamma, \, r] \equiv n'(\gamma) \, V'$, and $A = 2/3$ or $1$
for 2D or 3D expansion, respectively (see, e.g., Stawarz et al. 2004). 
Thus, the observed monochromatic --- e.g., X-ray --- inverse-Compton 
luminosity is evolving along the adiabatically expanding (in 2D, the
case considered hereafter for illustrative purposes) and non-decelerating 
jet as $L_X(r) \propto r^{-2 \, (p-1) / 3}$, where we assumed a power-law 
form for the electron energy distribution $n'(\gamma) \propto \gamma^{-p}$. 

Meanwhile, changes of the observed synchrotron luminosity at some given 
observed frequency $\nu_{syn} \sim 3 \, e \, B \, \delta \, \gamma^2 / 
4 \, \pi \, m_e c$, where $B$ is the jet magnetic field intensity, can 
be found from the appropriate expression
\begin{equation}
[\nu L_{\nu}]_{syn} \sim \delta^4 \, c \, \sigma_T \, V' \, U'_B \, 
\left[\gamma^3 \, n'(\gamma) \right]_{\gamma = \sqrt{4 \, \pi \, 
m_e c \, \nu_{syn} / 3 \, e \, B \, \delta}} \quad ,
\end{equation}
\noindent
where $U'_B = B^2 / 8 \pi$ is the jet comoving magnetic field energy
density. Thus, the observed monochromatic synchrotron --- e.g., radio
--- luminosity scales as $L_R(r) \propto r^{-2 \, (p-1)/3} \,
B^{(p+1)/2}$, since magnetic field intensity does change within
adiabatically expanding outflow, $B = B(r)$. As a result, the ratio of
the observed inverse-Compton to synchrotron luminosities at chosen
observed frequencies (e.g., $1$ keV and $1.4$ GHz) reads as $[L_X /
L_R](r) \propto B^{-(p+1)/2}$, assuming that the shape of the electron
energy distribution is roughly the same all along the jet (which is a
good approximation for the value $p \sim 3 \pm 0.5$ suggested by the
observed range of the X-ray and low-frequency radio indices in the PKS
1127-145 jet). Note that the $L_X/L_R$ ratio is then completely
described by the magnetic field evolution and the electron spectral
index so long as $\delta$ does not change down the jet.

It follows immediately from the above that in the case of adiabatic
expansion of a jet with conserved magnetic energy flux, both
inverse-Compton \emph{and} synchrotron luminosities decrease along the
outflow. For example, if the jet magnetic field consists predominantly
of a poloidal or toroidal component, the field is expected to scale
with a distance as $B(r) \propto r^{-2}$ or $\propto r^{-1}$,
respectively. The former (`poloidal') scaling cannot however hold
along the entire jet length, i.e. from the jet base (a few to several
Schwartzschild radii from the nucleus) up to the jet termination point
(about $300$ kpc from the active center). Indeed, taking the maximum
magnetic field intensity $B(10^{-3} \, {\rm pc}) \sim 10^4$ G at the
jet base (as appropriate for ten Schwartzschild radii from the central
black hole with mass $\sim 10^9 \, M_{\odot}$) it would imply
unrealistically low intensities on hundred kpc scales, $B(100 \, {\rm
kpc}) \sim 10^{-12}$~G (see recent discussion in Sikora et al. 2005
regarding the magnetic field in quasar jets). Therefore we consider
only the $B(r) \propto r^{-1}$ case (see in this context Tavecchio et
al. 2004), in which the X-ray (inverse-Compton) luminosity scales with
distance as $L_X(r) \propto r^{-4/3}$ while the radio (synchrotron)
one as $L_R(r) \propto r^{-10/3}$ for the assumed $p=3$. This scaling
of luminosities indicates that in the framework of the one-zone IC/CMB
model no adiabatic expansion of the PKS 1127-145 jet is possible,
since otherwise the expected attenuation of the X-ray and radio fluxes
would be inconsistent with the data. Indeed, taking the positions of
knots O and C relative to the core as $r_O = 7.3''$ and $r_C =
28.5''$, respectively (see Table 2), one should expect
$L_X(r_C)/L_X(r_O) \sim 0.2$, and $L_R(r_C)/L_R(r_O) \sim
0.01$. Meanwhile, the observations indicate that although the $1$ keV
luminosity decreases by about an order of magnitude between knots O
and C, the $1.4$~GHz luminosity increases by as much as a factor
of $\sim 10$ instead of decreasing be a factor of 100..

Given the above conclusion we now consider an efficient confinement of the 
outflow, i.e. no jet expansion, which would give the slowest attenuation
of the synchrotron (radio) luminosity. We also allow for the jet deceleration,
$\delta = \delta(r)$, and non-adiabatic changes of the jet magnetic field 
due to the related dissipation of the jet kinetic energy. The expressions 
(1) and (2) give then $L_X(r) \propto \delta^{(p+3)}$ and $L_R(r) \propto 
\delta^{(p+5)/2} \, B^{(p+1)/2}$.
In this case, an increase of the radio luminosity could be obtained if
the jet magnetic field is amplified. For instance, for a choice of
$\delta(r_C)/\delta(r_O) \sim 0.7$ and $B(r_C)/B(r_O) \sim 7$ the
X-ray luminosity decreases as $L_X(r_C)/L_X(r_O) \sim 0.1$, while the
radio luminosity increases like $L_R(r_C)/L_R(r_O) \sim 10$, resulting
in (required by the data) a factor of $\sim 0.01$ change in the
X-ray-to-radio luminosity ratio. However, such changes would imply a
significant amplification of the jet magnetic field energy density,
since $U'_B \propto B^2$ so that $[B(r_C) / B(r_O)]^2 \sim 50$. If,
therefore, the energy equipartition between the jet magnetic field and
the radiating electrons is valid for knot O, $U'_B(r_O) \sim
U'_e(r_O)$, then for knot C the equipartition would have to be broken,
$U'_B(r_C) \gg U'_e(r_C)$. In order to keep the energy equipartition,
one would have to assume that the normalization of the electron energy
distribution also increases along the jet just like $U'_B$, giving
$L_X(r) \propto \delta^{(p+3)} \, B^2$ and $L_R(r) \propto
\delta^{(p+5)/2} \, B^{(p+5)/2}$, i.e. that the amount of the jet
energy dissipated to the radiating particles is always equal to the
one transferred to the jet magnetic field. This would imply then
slightly more efficient deceleration and slightly less efficient
magnetic field amplification, namely $\delta(r_C)/\delta(r_O) \sim
0.4$ and $B(r_C)/B(r_O) \sim 4$, to reproduce the observed luminosity
profiles.

Sambruna et al. (2006) fit the `one-zone' IC/CMB model to two X-ray
jets, in the quasars 1136-135 and 1150+497, with similar jet luminosity
trends to that observed in PKS~1127-145, and obtained exactly the same
change in IC/CMB model parameters as calculated above, i.e. lowering
$\delta$ with increase in the B field and in the normalization of the
electron distribution along the jet (see their Fig. 12). As shown above, 
this is in fact a natural consequence of the equipartition requirement in 
the IC/CMB model if the decreasing (along the outflow) X-ray luminosity
anticorrelates with the radio one. Tavecchio et al (2006) proposed
that such changes in the jet parameters and the resulting radio and
X-ray profiles are due to the entrainment of the intergalactic medium
by the large-scale jet, but as shown by Hardcastle (2006) the entrained 
masses required are implausibly high in some cases.

Yet, the observed changes of the X-ray and radio spectral indices
along the PKS~1127-145 jet pose problems to the Inverse-Compton scenario.
Indeed, in the framework of the IC/CMB model electrons emitting the
observed $1$ keV photons have Lorentz factors $\gamma_X
\approx 10^3 / \delta$, while the electrons emitting the observed
$1.4$ GHz synchrotron radiation have Lorentz factors $\gamma_R \approx
10^4 / \sqrt{B_{-5} \, \delta}$, where $B_{-5} \equiv B / 10^{-5}$ G. 
Thus, for any Doppler factor $\delta \sim few$ and a magnetic field
close to the equipartition value $B_{-5} \lesssim 1$ (see
Siemiginowska et al. 2002, Kataoka \& Stawarz 2005) one has $\gamma_X
< \gamma_R$. This means that radiative cooling, if being a
dominant factor shaping the spectral indices' profiles along the PKS
1127-145 jet, would be more pronounced in radio frequencies than in
the X-rays. In other words, the radio continuum should have to steepen
more significantly with distance than the X-ray continuum.  In fact
the observed behaviour is just the opposite: the radio spectral
indices decrease along the jet, while the X-ray spectral index eventually
increases. Which is even more surprising is that in the outer portion
of the jet the X-ray spectral index is larger than the radio one, what
--- in the framework of the IC/CMB model --- would imply that the
energy distribution of the radiating electrons unexpectedly
\emph{steepens} towards the lower energies.

Thus, we conclude that the simplest one-zone IC/CMB model is not
consistent with the observed morphological and spectral properties of
the PKS 1127-145 radio/X-ray jet unless arbitrary additional
assumptions are invoked. In any case, its adiabatic version can be
completely rejected.

\subsection{The `One-Zone' Synchrotron Hypothesis}

Because of the extremely large jet lengths and extremely short radiative
cooling timescales of electrons with energies required for the synchrotron
X-ray emission (Lorentz factors $\gamma_X \approx 10^8 / \sqrt{B_{-5}
\, \delta}$) one obvious requirement of the synchrotron hypothesis
is continuous acceleration of the high-energy electrons along the
jet. The simplest version of the `one-zone' synchrotron model would be
therefore a scenario in which some unspecified acceleration mechanism
is acting continuously and similarly within the entire uniform jet
volume, producing and \emph{maintaining} a uniform electron energy
distribution with the steep power law plus flat spectrum component at
high energies.  Such an energy spectrum could be characterized in a
zero-order approximation by two power-laws with indices $p_1$ and $p_2
< p_1$ at low and high electron energies, respectively. As in the case
of the inverse Compton hypothesis let us check if such a simple model
can account for the observed properties of the PKS~1127-145 jet.

Just like in the previous subsection, one can conclude that no
adiabatic expansion of the jet is possible, since the radio luminosity
would decrease along the jet much faster than the X-ray. 
An efficient confinement of the jet and non-negligible dissipation
of the bulk kinetic energy give a scaling of the high- and low-energy
synchrotron luminosities 
$L_X(r) \propto 
\delta^{(p_2+5)/2} \, B^{(p_2+1)/2}$ and $L_R(r) \propto 
\delta^{(p_1+5)/2} \, B^{(p_1+1)/2}$, and thus 
$[L_X / L_R](r) \propto \left(\delta \, B\right)^{(p_2-p_1)/2}$.
Lets take the illustrative spectral indices of the electron energy
distribution close to $p_1 = 3$ (representing the low-frequency radio
continuum) and $p_2 \leq 2$ (representing freshly accelerated,
i.e. uncooled X-ray emitting electrons). We note that for example 
relatively well understood stochastic acceleration of 
particles due to resonant scattering on turbulent MHD waves characterized 
by the energy spectrum $W(k) \propto k^{-q}$ is expected to result in 
formation of the steady state particle energy distribution $n_e(\gamma) 
\propto \gamma^{-p}$ where $p=q-1$ (for particle energies below the 
maximum energy defined by the radiative loss timescale; see the 
discussion in Kataoka et al. 2006). Thus, for the typical turbulence 
energy index $1 \leq q \leq 2$ one should indeed expect particle spectra with 
$0 \leq p \leq 1$, i.e. much flatter than the ones claimed by the shock 
scenarios ($p \sim 2$).

It follows immediately that the agreement between the predicted and
the observed luminosity profiles along the PKS~1127-145 jet is
possible only if the high-energy spectral component of the radiating
electrons is very flat, and if in addition the magnetic field is
strongly amplified along the jet. Namely, the decrease of the X-ray
luminosity by a factor of $0.1$ and the increase of the radio
luminosity by a factor of $10$ would be possible for $p_2 = 0$ only if
$\delta(r_C)/\delta(r_O)
\gtrsim 0.1$ and $B(r_C)/B(r_O) \gtrsim 100$, while $p_2 > 0$ would 
require even more drastic magnetic field amplification. These changes
of the jet parameters are worrisome, simply because it is difficult to
isolate the appropriate mechanisms causing such a strong jet
deceleration and extremely effective conversion of the jet kinetic
energy to the magnetic one (see again comments by Hardcastle 2006 on
the entrainment process on large scales). Although it can be argued
that \emph{some} magnetic field amplification and \emph{some} bulk
deceleration are taking place along the extragalactic large-scale jets
anyway, as suggested by the observed in many cases 
the increase in the radio surface brightness
and decrease in the jet-couterjet brightness asymmetry away from a
core, the particular model parameters obtained
above are rather too extreme and should be considered implausible.

On the other hand, the assumption about the uniform electron energy
distribution within the entire jet is not a natural nor a realistic
one for the synchrotron/continuous acceleration scenario. In fact, if
the high-energy flat-spectrum (pile-up) component in the electron
distribution is produced, it results most probably from turbulent
acceleration processes. Thus, the normalization, the effective
spectral index, and the cut-off energy of such a pile-up component
should depend on the local plasma conditions at a given distance along
the jet (magnetic field intensity, turbulence parameters, jet velocity
structure, etc.). These factors can influence significantly model
predictions when compared to the simplest scenario discussed above
(which can hardly explain the observed radio/X-ray flux profiles along
the PKS~1127-145 jet). Discussing all these effects would require much
deeper insight into the microscopic processes in relativistic
magnetized plasma, which are still hardly known. We can however
consider briefly one simple possibility and discuss the consequences
of the assumption that the normalization of the high-energy flat
spectrum electron component producing synchrotron X-rays depends on
the kinematic parameters of the jet. Since the jet shear boundary
layer is the site of choice for generation of the high energy
electrons (Stawarz \& Ostrowski 2002), one could speculate that the
efficiency of the particle acceleration process (and thus the
normalization of the electron distribution $K_2$, where $n_e(\gamma) =
K_2 \, \gamma^{-p_2}$ at high energies) is somehow proportional to the
velocity shear across the relativistic outflow, $d \Gamma / d R \sim
\Gamma /R_j$ (where $R$ is the radial jet dimension and $R_j$ is the
jet radius), at any distance from the core. That is to say, for the
confined cylindrical jet ($R_j = const$) and a constant jet viewing
angle, we introduce a simple scaling $K_2
\propto
\delta^{m}$ with the power-law coefficient being $m>0$ for generality. In such 
a case, with $p_2 = 0$ and $p_1 = 3$ considered previously, the
synchrotron X-ray and radio luminosities should go like $L_X \propto
\delta^{(2m+5)/2} \, B^{1/2}$ and $L_R \propto \delta^{4} \,
B^{2}$. Taking therefore $m=5$ as an arbitrary illustrative value, one
can find that moderate changes $\delta(r_C)/\delta(r_O)
\sim 0.6$ and $B(r_C)/B(r_O) \sim 8$ reproduces well the luminosity 
profiles observed along PKS 1127-145 jet. These are then not
implausible anymore.

However, the observed changes of the X-ray and radio spectral indices
are again problematic for the simple synchrotron model discussed
here. While it is true that one should expect some ageing of the X-ray
continuum, it is not clear why the observed increase of the X-ray
spectral index is so gradual, and in addition accompanied with a
flattening of the radio spectrum. If one assumes that the freshly
accelerated high-energy part of the electron energy distribution is
indeed very flat ($p_2 < 2$) then the spectral index of the constantly
injected radiatively cooled (by the standard synchrotron-type energy
losses) high-energy particles should be equal to $\alpha = 0.5$. This
is quite close to the X-ray spectral index observed within the inner
jet. Meanwhile, in the outer jet the X-ray spectral index steepens,
suggesting either a significant change in the initial power-law slope
of the electron pile-up component (down to $p_2 \gtrsim 2$), or
significant differences in the cooling mechanism in the inner and
outer jet.  Whatever the case, it is hard to understand why in the
inner jet, where the postulated pile-up emission would be relative
flat (and hence the required acceleration process is especially
efficient), the observed synchrotron continuum at high radio
frequencies is very steep.

To summarize, we conclude that also the simplest one-emitting zone
version of the synchrotron model with uniform (though non-standard,
i.e. concave) electron energy spectrum, cannot easily explain the 
observed morphological and spectral properties of the PKS 1127-145 
radio/X-ray jet.

\subsection{A Two-Component Model: the Jet and a Sheath.}

A natural solution to the problems encountered by the homogeneous 
one zone models is to assume that the observed radio and X-ray radiative
components are produced in separate regions, possibly by separate 
and different acceleration processes. Here, we suggest that
indeed the detected X-ray photons originate within the {\it proper
jet} flow, while the radio photons come from a {\it sheath} --- a 
slow moving radial {\it extension} of the jet boundary (mixing)
layer. Such a possibility is in fact not novel, since it is a
variation of the stratified jet scenario (see De Young 1986,
Komissarov 1990, Laing 1996, Ostrowski 2000, discussing different
aspects of this issue), noted before in the particular case of 
the {\it Chandra} quasar jets by Hardcastle (2006). We emphasize, 
that this {\it proper jet-sheath} model is not exactly the same 
as the {\it spine-shear layer}. The {\it sheath} here is associated 
with some extension of the jet boundary (similarly to 3C~273 jet/cocoon 
system; see Bahcall et al. 1995), while the {\it proper jet} means a 
fast spine together with the shear layer.

In a framework of the two-component jet model, all the information
regarding the broad-band jet emission is restricted to the X-ray
band. In other words, there are not many additional constraints which could
help to distinguish between the synchrotron and inverse-Compton origin
of the observed X-ray keV photons --- both these possibilities have to
be kept in mind. However, in an analogy to the X-ray jet in 3C 273,
where almost the same gradual decrease of the X-ray flux and
steepening of the X-ray continuum along the jet is observed, one could
argue that also in the case of PKS~1127-145 the X-ray radiation is
most probably due to the synchrotron emission of the high energy
electrons accelerated continuously within the jet volume (see a
discussion in Uchiyama et al. 2006, Jester et al. 2006, and references
therein).

Note also, that assuming  a radio-to-X-ray continuum characterized by a
single power-law with a slope $\alpha_{R-X}
\lesssim 0.6$ (corresponding to the particle spectral index $p \lesssim 2.2$), 
the expected radio emission of the proper jet would be always below
the observed radio emission from the PKS~1127-145 jet regions, since
all the radio-to-X-ray spectral indices shown in figure 10 are $>
0.6$. The same is true for the optical emission, as the optical upper
limits available for the knots B and C (Siemiginowska et al. 2002)
imply optical-to-X-ray power-law slopes $< 1.0$. That is not to say
that the jet broad-band emission is in  reality  a single power-law
form. Indeed, it may be characterized by a more complicated spectral
shape for which however we do not have any strong evidence.

By definition, a two-component model doubles the  number of free
parameters, making model predictions more flexible when confronted
with the data. Lets therefore briefly discuss whether the
observational X-ray/radio constraints (i)-(iv) listed at the beginning
of this section could be explained in a natural way in a framework of
such a scenario.

\subsubsection{The Proper Jet}

The observed smooth decrease of the X-ray flux, when unconnected with
the increase of the radio one, is in fact expected as a result of the
decrease in the inverse-Compton or synchrotron luminosities along
(even extremely slowly) decelerating and/or expanding jet. These
luminosities (see equations 1 and 2 above) can be expressed as
\begin{equation}
L_{ic} = \int [L_{\nu}]_{ic} \, d\nu \sim \delta^6 \, \Gamma^{-2} \, {c \, \sigma_T \over m_e c^2} \, 
V' \, U'_{cmb} \, {\langle\gamma^2\rangle \over \langle\gamma\rangle} \,
U'_e \quad ,
\end{equation}
\noindent
and
\begin{equation}
L_{syn} = \int [L_{\nu}]_{syn} \, d\nu \sim \delta^4 \, {c \, \sigma_T \over m_e c^2} \, V' \, U'_B \, 
{\langle\gamma^2\rangle \over \langle\gamma\rangle} \, U'_e \quad ,
\end{equation}
\noindent
respectively, where $U'_{cmb} = 1.2 \times 10^{-11} \, \Gamma^2$ erg
cm$^{-3}$ at the redshift of PKS 1127-145, where the factor $\langle\gamma^2\rangle / 
\langle\gamma \rangle \equiv \int \, \gamma^{2-p} \, d\gamma / \int \gamma^{1-p} 
\, d\gamma$ characterizes the electron energy spectrum $n'(\gamma) \propto 
\gamma^{-p}$ with $p \sim 3$ considered hereafter, and 
$V' \sim \pi R^2 \, l'$ is the comoving emitting volume 
of some cylindrical portion of the jet with the longitudinal and radial 
(with respect to the jet axis) dimensions $l'$ and $R$, respectively. 
For example, assuming the energy equipartition $U'_e(r) \sim 
U'_B(r)$ along the jet, an attenuation of the jet magnetic field $B(r)
\propto r^{-k}$ (where $k > 0$), and a constant jet's (small) opening angle 
$\vartheta \sim R / r$, one obtains the synchrotron luminosity profile
$L_{syn}(r) \propto r^{2-4 \, k} \, \delta^4(r)$, if only the broad-band 
spectral shape of the emitting electrons (i.e., factor $\langle\gamma^2\rangle /
\langle\gamma\rangle$) is roughly constant along the jet. Thus, the observed 
decrease of the X-ray luminosity, by a factor of $14$ between knots I
and C (see figure 9), requires $k \approx 0.8$ for the negligible
changes in the jet Doppler factors. This, in turn, implies a decrease
of the jet magnetic field by a factor of $\approx 0.2$ along the jet,
which is the most extreme change needed, i.e. a lower limit, since
some non-negligible (though rather small) decrease of the jet Doppler
factor should be expected. Indeed, if the jet Doppler factor decreases
by only a factor of $2$ between knots I and C, then no decrease of the
jet magnetic field intensity is needed to explain the observed
decrease of the X-ray (i.e., synchrotron) luminosity along the PKS
1127-145 jet. This is much more comfortable situation than a
significant amplification required by one-zone models discussed above.

Similarly, in the case of the inverse-Compton X-ray luminosity
assuming the radiating electrons-magnetic field energy equipartition
$U'_e(r) \sim U'_B(r) \propto r^{-2 \, k}$, one gets $L_{ic}(r) 
\propto r^{2-2 \, k} \, \delta^6(r)$,
leading to the required $k \approx 1.65$ for the negligible changes in
the jet Doppler factor between knots I and C. This would imply
a significantly larger decrease of the jet magnetic field intensity than
before (by a factor of $0.03$) but, again, any small decrease of
$\delta(r)$ along the jet can adjust required parameters.

We now check if the observed gradual increase in the jet X-ray
spectral index (i.e. when decoupled from flattening of the radio
continuum) can be incorporated into the synchrotron and the
inverse-Compton scenarios, by means of adjusting the appropriate
radiative losses and dynamical time-scales.  We note, that at
the redshift of PKS~1127-145, the comoving energy density of the jet
magnetic field dominates over the energy density of the CMB radiation
as measured in the jet rest frame, $U'_B > U'_{cmb}$, if only $B > 20
\, \Gamma$ $\mu$G. This condition is difficult to fulfill, since in
the case of quasar large-scale jets we expect $\Gamma >1$ and $B< 10$
$\mu$G. Thus, radiative cooling of the jet electrons in PKS~1127-145
is expected to be mainly due to inverse-comptonization of the CMB
photons.

The transition between slow- and fast-cooling regimes for the X-ray
emitting electrons occurs at the distance along the jet, $r_{br}$,
where the dynamical time-scale $t'_{dyn} = r/ c \, \Gamma$ equals the
radiative losses time-scale $t'_{rad} = 3 \, m_{\rm e} c^2 / 4 \,
\sigma_{\rm T} c \, \gamma_X \, U'_{cmb}$. Here $U'_{cmb} = (4/3) \,
U_{cmb} \, (1+z)^4
\, \Gamma^2$, and the $z=0$ value of the CMB energy density is $U_{cmb} \approx
4 \times 10^{-13}$ erg cm$^{-3}$. With $\gamma_X = (\nu_X / \nu_{cmb}
\, \delta^2)^{1/2}$, the observed X-ray photon energy $h \, \nu_X = 1$ keV, and the
energy of the CMB photons $h \, \nu_{cmb} = 1$ meV at the redshift $z=0$, one can find
\begin{equation}
r_{br} \approx {9 \, m_{\rm e} c^2 \, \delta \, \nu_{cmb}^{1/2} \over 16 \, \sigma_{\rm T} \, 
U_{cmb} \, (1+z)^4 \, \Gamma \, \nu_X^{1/2}} \sim
25 \, \delta \, \Gamma^{-1} \, {\rm Mpc} \, .
\end{equation}
\noindent
This, when compared to the projected distance of knot B at which the
break in the X-ray spectrum is observed, $r_B \sim 18'' \sim 150$ kpc,
is extremely large. In fact, it can be found that the deprojected
distance of knot B equals the expected break radius only for
unrealistic jet parameters. Namely, $r_B/\sin \theta \sim r_{br}$
(where $\theta$ is the jet viewing angle) requires large bulk Lorentz
factors of the jet and very large jet inclinations, $45$~deg $< \theta
<$ $90$~deg for $13 < \Gamma < 20$, leading to the energetic problems.
This means that the IC/CMB model cannot explain the observed
steepening of the X-ray continuum along the PKS~1127-145 jet in terms
of the spectral ageing of the low-energy electrons. Instead, an
\emph{`intrinsic'} change of the electron spectral shape (i.e. not due
to the radiative cooling but particle acceleration processes) has to
be invoked to explain observations, what is also the case for the
synchrotron scenario (for which analogously evaluated $r_{br}$ is
much smaller than $r_B$).

\subsubsection{A Sheath}

Now we turn our attention to the jet sheath which, by assumption,
dominates the observed radio emission of the PKS 1127-145 jet. Here we
investigate if the sheath hypothesis can explain in a simple way the
observed increase of the radio flux along the jet accompanied by a
flattening of the radio continuum.  Note first, that a very similar
spectral (radio) behaviour is in general observed in all lobes of FR
II radio galaxies.  In particular, a steepening and an attenuation of
the radio emission away from the jet termination point in the
direction of the nucleus observed in FR II sources is widely
interpreted as a result of the spectral ageing of the radio-emitting
plasma (see Kaiser 2000 and references therein). In the case of the
PKS~1127-145, however, we propose that the observed radio emission
originates from some radial extension of the outflow rather than from
the radio lobe (i.e. plasma backflowing from the jet termination hot
spot).  Thus the question we should answer is why the inner jet (knots
I--A) produces a very weak and steep-spectrum radio sheath, while the
outer jet (knots B and C) a relatively bright and flat spectrum
one. (On the other hand, a strict distinction between a jet sheath and
a lobe may be quite artificial, especially in the outer portions of
the jet.)

We propose that the reason for the observed PKS~1127-145 jet behaviour
is the difference in the dominant cooling process of the
radio-emitting sheath's electrons: a radiative cooling within the
inner sheath (surrounding the inner jet), and an adiabatic cooling
within the outer one (corresponding to the outer jet). As is well
known, the frequency independent adiabatic losses cannot significantly
affect the spectral shape of the synchrotron emission, while the
frequency-dependent radiative losses result in a spectral steepening
above a given frequency. Such a steepening may be significant
(resulting even in spectral cut-offs at low radio frequencies),
depending on the detailed evolution of the radiating particles, which
in turn depends on the structure of the magnetic field in the emission
region. This could possibly explain a flat radio spectrum outer sheath
and steep spectrum outer sheath. In order to investigate this idea in
more detail, we assume a non-relativistic bulk velocity of the sheath
plasma, and inefficient particle acceleration processes acting
thereby. This is rather for illustrative purposes only and an
order-of-magnitude estimate, since in the observer's rest-frame the
sheath material may be still moving relativistically, if only the
inertia of the proper jet is big enough to ensure a relativistic
advance velocity of the jet's head. In fact, this may be the case of
PKS~1127-145, which can also help to understand an apparent lack of
the counter-sheath on the other side of the nucleus (just like in
3C~273 as discussed in Stawarz 2004).

The radiative (synchrotron and IC/CMB at $z=1.18$) cooling time scale
for the electrons emitting synchrotron photons at frequency $\nu$ can
be expressed as
\begin{equation}
t_{rad} \sim \min \left(10 \, B_{-5}^{-3/2}, \, 5 \, B_{-5}^{1/2} \right)
\, \nu_{10}^{-1/2} \quad {\rm Myrs} \quad ,
\end{equation}
\noindent
where `min' indicates a minimum of two values, $\nu_{10} \equiv \nu / 
10^{10}$ Hz and, as before, $B_{-5} \equiv B/10^{-5}$ G. Meanwhile, 
the time scale for the adiabatic losses can be approximated as
\begin{equation}
t_{ad} \sim {R \over dR/dt} \sim 0.03 \, R_{10} \, \beta_{ad}^{-1} \quad {\rm Myrs} \quad ,
\end{equation}
\noindent
where $R_{10} \equiv R / 10$ kpc, and we assumed that the sheath
expansion velocity is roughly constant at a given distance from the
nucleus, $c \beta_{ad} = dR / dt \sim const$. Our model requires then
$t_{rad} < t_{ad}$ for the inner sheath, and $t_{rad} > t_{ad}$ for
the outer one. As argued below, this may be in fact expected.

Assuming that the radio sheath in PKS~1127-145 is in a direct contact
with the gaseous environment of the intergalactic medium (IGM)
characterized by the particle number density $n_g$ and temperature
$T$, the velocity of the sideway expansion for the sheath can be found
as $\beta_{ad} \sim (p_{sh} / n_g \, m_p c^2)^{1/2}$, where $p_{sh}$
is the internal pressure of the radio emitting (ultrarelativistic)
plasma.  With the minimum power condition fulfilled within the sheath,
one has $p_{sh} \sim U_B$, and therefore $\beta_{ad} \sim 10^{-3} \,
B_{-5} \, n_{-3}^{-1/2}$, where $n_{-3} \equiv n_g /
10^{-3}$\,cm$^{-3}$. Now, for the inner sheath we assume pressure
balance $p_{sh} \sim p_g$, where $p_g = n_g \, k T $ is the thermal
pressure of IGM. This implies that the expansion of the inner sheath
is happening at the sound speed, $\beta_{ad} \sim (k T / m_p
c^2)^{1/2}$, i.e., $\beta_{ad}
\sim 10^{-3}$ for the expected $T \sim 10^7$\,K,
 and that the magnetic field intensity
within the sheath, $B \sim (n_g \, k T \, 8 \pi)^{1/2}$, is simply
$B_{-5} \sim 0.6 \, n_{-3}^{1/2}$. Interestingly, for the expected
$n_{-3} \lesssim 1$ on $10-100$ kpc distances from the center of the
typical galaxy group environment (see, e.g., Mathews \& Brighenti
2003), coinciding by assumption with the core of PKS 1127-145 radio
source, our simple model gives very realistic value $B_{-5} \lesssim
1$, and thus the radiative lifetime $t_{rad} \lesssim 10$ Myrs for the
electrons emitting $\geq 1$ GHz synchrotron photons. This timescale,
as required, is then shorter than the expansion timescale $t_{ad} \sim
30 \, R_{10}$ Myrs. Moreover, the obtained radiative cooling is
roughly comparable to the jet lifetime at the position of knot A,
which can be crudely evaluated as $t_{life} \sim 1-10$ Myrs for the
jet advance velocity $\sim (0.1-1) \, c$ and the assumed jet viewing
angle $\sim 15\deg$. Thus, a very weak (and steep-spectrum) radio
emission at the position of knot A may be explained by the sheath
hypothesis.

Meanwhile, further away from the core, at the position of the outer jet, 
a decreased pressure of the ambient medium is likely to result in a more 
rapid expansion of the sheath. Namely, anticipating relatively steep 
profile of the IGM density $n_g \propto r^{-\zeta}$ with $\zeta \geq 2$, 
one can expect thermal medium surrounding the outer sheath being as rarified 
as $n_{-3} \lesssim 0.1$. In this case, taking $B_{-5} \sim 1$ as before, one 
obtains the adiabatic cooling timescale $t_{ad} < 10 \, R_{10}$ Myrs, likely
shorter than (or at least comparable with) the radiative timescale for the 
electrons emitting $\leq 10$ GHz synchrotron photons, again as required by 
the model. Thus, dominant adiabatic cooling can indeed result in the 
flat-spectrum radio emission at the position of knots B and C. Interestingly,
the expansion velocity of the outer sheath in a framework of the proposed
scenario is expected to be larger than the sound speed in IGM, $\beta_{ad} 
> 10^{-3}$. This can possibly result in formation of a weak shock in the
thermal medium being in the closest vicinity of the radio structure. 
Further investigation of this issue, involving analysis of the X-ray 
environment of PKS~1127-145 radio source, is in preparation.

\subsection{Modulated Jet Activity}

The bolometric luminosity of the PKS~1127-145 nucleus, based 
on the observed optical-UV isotropic continuum, is about $L_{nuc} 
\sim 8 \times 10^{46}$erg~sec$^{-1}$ (B\l a\.zejowski et al. 2004). 
The black hole mass determined by Fan \& Cao (2004) is about $M_{BH} \sim 7 
\times 10^8$~M$_{\odot}$ which gives an Eddington luminosity of $L_{Edd} 
\sim 9 \times 10^{46}$ erg~sec$^{-1}$. Thus, PKS~1127-145 seems to
accrete close to the Eddington limit, since $L_{nuc} / L_{Edd} 
\sim 1$. During the jet activity epoch the quasar would then 
provide a total energy of
\begin{equation}
E_{tot} \sim 2.5 \times 10^{58} \, \left({\eta \over 0.01}\right) \, 
\left({t_{life} \over {\rm Myr}}\right) \quad {\rm ergs} \, , 
\end{equation}
\noindent
where $\eta \equiv L_{jet}/L_{nuc}$ is the efficiency factor of converting
the accretion power into the kinetic power carried with the jet. 
Because the  $>200$ kpc projected length of the PKS~1127-145 jet implies
$t_{life} > 3$ Myrs (with the speed of light providing an upper limit
for the jet advanced velocity, and with the illustrative jet viewing angle of 
$15$~deg), the total energy transported by the jet 
is $E_{tot} > 10^{59}$ ergs, assuming continuous jet activity (i.e. constant 
kinetic power of the outflow during the entire activity epoch)

In this context we raise an issue on the nature of the knots in
PKS~1127-145: can they be considered as the extended shock waves
formed within a more or less continuous jet outflow? In our opinion
not, since the projected linear sizes of these knots are uncomfortably
large ($> 10$ kpc). Instead, we believe that it is much more natural
to consider knots I, O, A, B, and C as representing the separate
portions of the jet produced by the active PKS~1127-145 core in
separate epochs of its higher activity (see a discussion in Stawarz et
al. 2004 and Siemiginowska et al 2002). Since all of the PKS 1127-145
knots are a few tens of kpc long, the postulated duration of a
continuous activity epoch would be roughly $\sim 10^5$ yrs. It is
therefore not surprising that this kind of intermittency is the most
pronounced in PKS~1127-145, possessing the longest X-ray jet known.

Interestingly, the radio core of PKS~1127-145 itself has been
classified as a Giga-Hertz-Peaked Spectrum (GPS) source (Stanghellini
et al. 1998), which are established to be young, i.e. $\lesssim 10^5$
yrs old, versions of powerful radio galaxies (see O'Dea
1998). However, as argued by B{\l}a\.zejowski et al. (2004) based on
the multiwavelength spectrum and the luminosity dominated by
$\gamma$-rays, the unresolved core (at least the component producing
its high-energy emission) seems to be related rather to the blazar
phenomenon (see in this context also Tornikoski et al. 2001). The
presence of $> 200$-kpc-long radio/X-ray jet additionally questions the
GPS nature of PKS 1127-145 quasar. On the other hand, in a framework
of the `modulated jet activity' scenario --- evidenced by
double-double radio galaxies (Schoenmakers et al. 1999) and postulated
by Reynolds \& Begelman (1997) to explain statistics of GPS sources
--- there may be no contradiction in the above. Namely, the blazar
core of PKS~1127-145 at the moment of the observation could be
considered as being again in the very active state, i.e., producing
another portion of the jet matter with the enhanced kinetic
luminosity. At this moment, the new-born portion of the jet resembles
the `classical' GPS object, but at some later time it will be observed
as an extended jet `knot' (similar to the knots I, O, A, B and C
discussed in this paper).

We note that signatures of intermittent AGN jet activity have been suggested 
also by many recent {\it Chandra} observation of X-ray clusters (Fabian et al
2003, McNamarra et al 2005). Typical timescales for this activity can
be determined from the cluster X-ray and radio morphology, and they
are within 10$^4 -10^8$~years depending on the source 
(see, e.g., Nulsen et al. 2005a, 2005b, Forman et al. 2005, 2006). 
A total energy of $10^{58}-10^{60}$~ergs contained within large
radio cavities in the most powerful clusters provides an estimate of
the total energy carried out by the jet and radio plasma, consistent
with our estimates for the quasar PKS~1127-145.

\section{Summary and Conclusions}

The new X-ray and radio data of the PKS~1127-145 jet do not provide an
unambiguous conclusion on the dominant jet X-ray emission process. It
is clear that simple synchrotron or inverse-Compton models do not
apply and more complex models need to be developed.  However, one
needs to remember that the above discussion is based on averaging
X-ray and radio morphologies that do not closely match and ignoring
any individual sub-structures that might be present in the jet.  As
noted at the beginning of the discussion the PKS~1127-145 jet is about
10 times longer than the 3C~273 jet, while characteristic timescales
for energy losses in both jets are similar. In addition, nearby
sources (e.g. M87 or Cen A) show that jets have smaller scale
structures that cannot be resolved in the higher redshift
sources. 

The X-ray and radio morphology and spectral properties of
the PKS 1127-145 jet are similar to the 3C 273 jet, indicating that the
3C~273 jet is not unusual but a representative case for other quasar
jets. The homogeneous one-zone (either IC/CMB or synchrotron) models
are not consistent with the PKS~1127-145 X-ray and radio data, and their 
adiabatic versions can be rejected in general. A possible solution 
is a model in which the radio emission is produced
within a jet sheath, while the X-rays come from the proper jet. This
implies possibly a much different spectral and spatial evolution of
the X-ray and radio spectra. In addition to this, we speculate 
that the jet intermittency also plays a role in shaping the observed
jet morphology in PKS 1127-145.

One needs to remember that the two highest redshift X-ray jets known
to date (GB\, 1508+5714, Siemiginowska et al. 2003; 1745+624, Cheung
et al. 2006) are faint and the resolution scale in X-rays is of order
$\sim 10$~kpc. Thus the conclusions about the dominant emission
processes are based on averaging over unknown jet structures.
Statistical analysis of large samples of high redshift jets may be the
only method available to study jet properties and their impact on the
environment in the early Universe.

\acknowledgments

The authors thank the anonymous referee for helpful comments.
This research is funded in part by NASA contract NAS8-39073.  Partial
support for this work was provided by the National Aeronautics and
Space Administration through Chandra Awards Number GO5-6113X,
GO4-5131X, GO5-6118X, GO5-6111X issued by the Chandra X-Ray
Observatory Center, which is operated by the Smithsonian Astrophysical
Observatory for and on behalf of NASA under contract NAS8-39073.
\L .~S. acknowledges support by MEiN through the research project 
1-P03D-003-29 in years 2005-2008, and by the ENIGMA Network through
the grant HPRN-CT-2002-00321.  The National Radio Astronomy
Observatory is operated by Associated Universities, Inc. under a
cooperative agreement with the National Science Foundation.

{}

\newpage


\begin{table}
\begin{scriptsize}
\caption{\small Radio Observations}
\begin{center}
\label{tab-radio}
{
\begin{tabular}{lccccccc}
\hline
\hline\noalign{\smallskip}
Program & Obs. Date &   Config. & Freq. &  Integ. & Beamsize &rms  &   Dynamic \\
&&&&                                        Time &   &&
Range \\
&&&                               (GHz) &     (hr) & & (mJy beam$^{-1}$) & \\
\hline
\\
AH730   & 2001 Feb 6  &    BnA  &   1.4   &  2.1  & 2.71'' $\times$ 1.62'' at PA=84.4 deg  &0.30 &   18,000 \\
        &             &         &   8.5   &  3.4  & 0.63'' $\times$ 0.37'' at PA=77 deg   &0.053 &  65,000 \\
        & 2001 Jun 18 &    CnB  &   4.9   &  2.6  & 2.77'' $\times$ 1.66'' at PA=84 deg  &0.12 & 33,000 \\
AC779   & 2005 May 10 &     B   &   8.5   & 6.25  & 0.7'' $\times$ 0.7''  &0.04 & 81,000 \\
\\
\hline
\end{tabular}
\\
Notes:
The beamsizes indicate the FWHM of the major and minor axes of the elliptical Gaussian restoring beam used
in the images (Figures~6, 7 \& 8).
}
\end{center}
\end{scriptsize}
\end{table}


\begin{table}
\begin{scriptsize}
\caption{Parameters of the Jet Features}
\begin{center}
\label{tab1}
\begin{tabular}{lcccccccccc}
\hline
\hline\noalign{\smallskip}
 &D$^a$ &Counts$^b$ & Net$^b$& $\alpha_{X}$ & F(0.5-2~keV)$^c$ & F(2-10~keV)$^b$ & S(1.4)$^d$ & S(5)$^d$&S(8.5)$^d$ & $\alpha_R^e$ \\
\hline
\\
I& 3.8&475$^{\pm22}$ & 323.6$^{\pm25.1}$  & 0.67$^{+0.11}_{-0.11}$ & 7.9$^{\pm0.6}$  & 16.8$^{\pm1.3}$ &  & & 2.2$^{\pm1.0}$ & \\
O& 7.3&118$^{\pm11}$ & 89.9$^{\pm12.1}$   & 0.69$^{+0.19}_{-0.18}$ & 2.2$^{\pm0.3}$  & 4.6$^{\pm0.6}$ & 2.5$^{\pm0.3}$ & $<0.45$ & $<0.2$ & $>1.35$ \\
A& 11.2&156$^{\pm12}$ & 132.7$^{\pm13.4}$ & 0.66$^{+0.15}_{-0.15}$ & 3.2$^{\pm0.3}$  & 6.8$^{\pm0.7}$ & 6.4$^{\pm0.9}$  & 1.2$^{\pm0.2}$ & $<0.27$ & 1.32$^{\pm0.17}$ \\
B& 18.6&87$^{\pm9}$   & 77.35$^{\pm9.8}$   & 1.0$^{+0.2}_{-0.2}$ & 1.98$^{\pm0.25}$ & 2.6$^{\pm0.3}$ & 43.2$^{\pm4.3}$ & 14.4$^{\pm1.4}$ &8.2$^{\pm0.8}$ & 0.91$^{\pm0.07}$ \\
C& 28.5 &40$^{\pm6}$   & 20.9$^{\pm7.7}$    & 1.2$^{+0.6}_{-0.5}$ & 0.57$^{\pm0.21}$ & 0.54$^{\pm0.20}$ & 53.2$^{\pm5.3}$ & 16.7$^{\pm1.7}$ & 11.8$^{\pm1.2}$ & 0.85$^{\pm 0.08}$ \\
\\
\hline
\end{tabular}

\noindent
Notes: $^a$ D - Distance from the core to the center of the box region
measured in arcsec.  $^b$ Total or Net Counts with energy within
0.3-8~keV $^c$ Flux in units of 10$^{-15}$ erg~cm$^{-2}$s$^{-1}$; Flux
errors based on the counts errors only.  $^d$ Flux density in mJy (at
a given in the parentheses radio frequency are in GHz), estimated for the
regions defined by the X-ray features (see Fig.\ref{jet-reg}). The rms
in the maps and the beam size is given in Table~\ref{tab-radio}
The size of the regions assumed for extracting the spectra:
I:3.4''x1.98'', O:3.1''x2.36'',A:5.12''x2.64'', B:4.41''x2.13'',
C:7.03''x5.63''. $^e$ Radio spectral index measured between 1.4 and
8.5~GHz.
\end{center}

\end{scriptsize}
\end{table}

\newpage


\begin{figure}
\epsscale{0.85}
\caption{Figure available at 
{\tt http://hea-www.harvard.edu/$\sim$aneta/PKS1127\_2006/}
ACIS-S exposure corrected image (E=0.3-7~keV) which has been smoothed
with the Gaussian kernel (1$\sigma$=0.615~arcsec). The quasar core has
been excluded from the image within a circular region with 3~arcsec
radius. The knots are marked with labels along the jet. The ``kink''
region in the jet is marked by an arrow.  The direction of the CCD
readout is indicated by an arrow on the right. The color scale is
marked on a color bar with units of 8.13
photons~cm$^{-2}$~s$^{-1}$~arcsec$^2$.  The sky coordinates are in
J2000. North is up and East is to the left.}
\label{smooth}
\end{figure}

\begin{figure}
\epsscale{0.75}
\plottwo{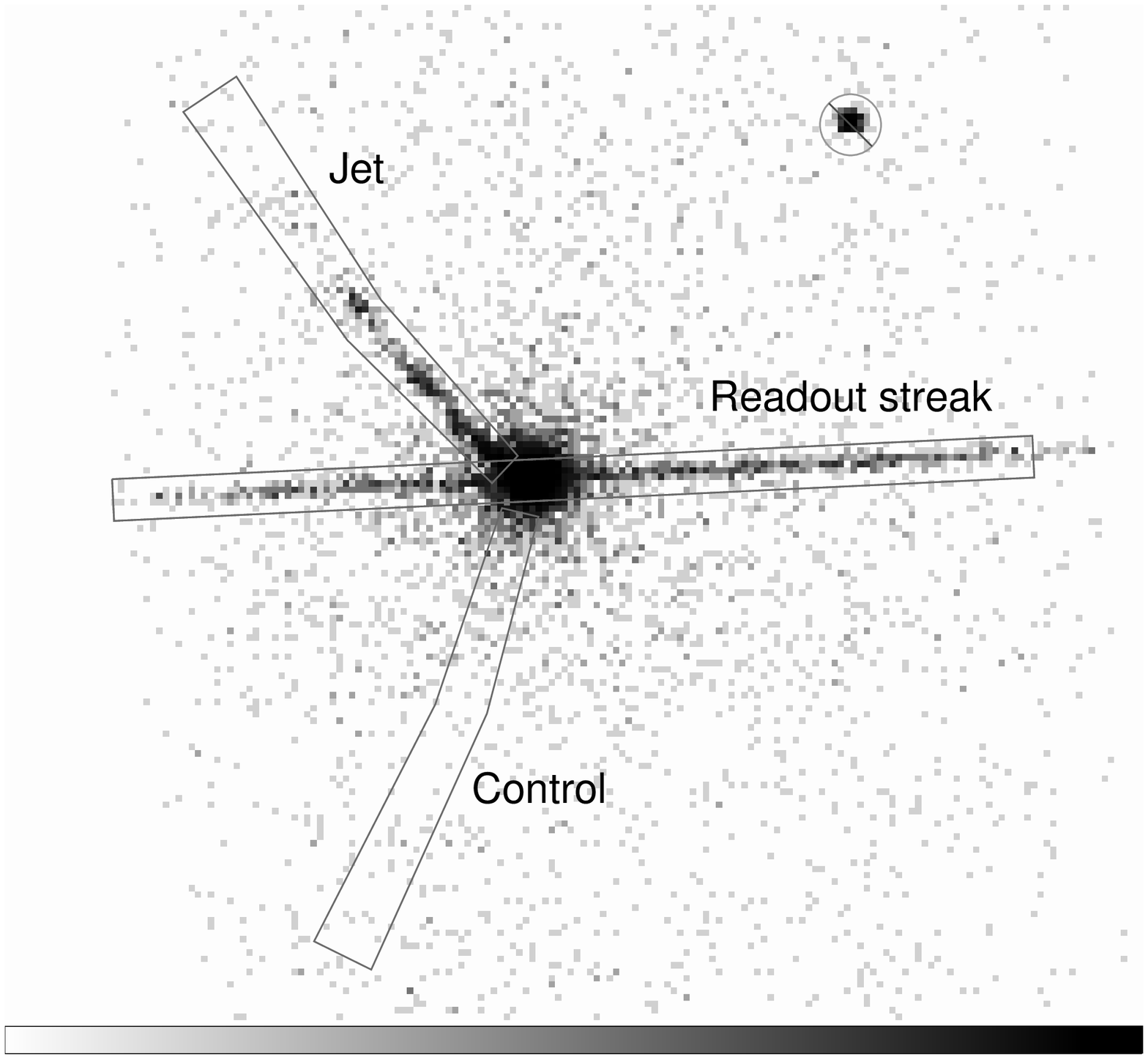}{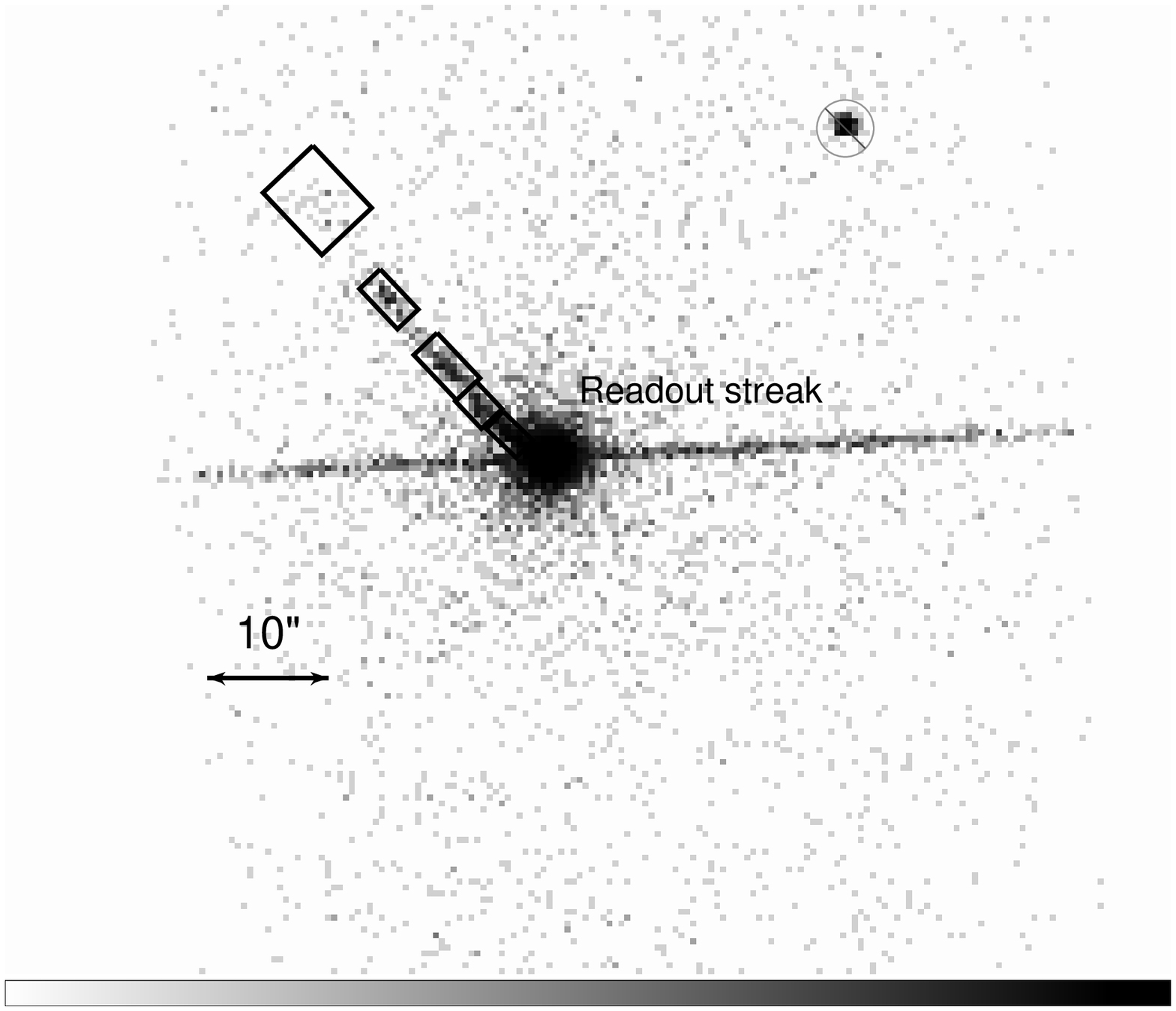}
\caption{(a) Regions assumed for the extraction of the jet 
profile and the control
profile are shown in the raw ACIS-S image (E=0.3-7~keV). The readout
streak regions and the point source to the north-west marked in the
image have been excluded from the analysis.  (b) ACIS-S raw image data
overlayed with the spectral regions (I,O,A,B,C - counting away from
the quasar along the jet) assumed for the extraction of the jet X-ray
spectra.  North is up and East is left. The readout streak is marked
with label and it is visible on both East and West sides from the
quasar core. Pixel size is set to half of the original ACIS pixel, and
it is equal to 0.246~arcsec. The 10~arcsec scale is marked with an
arrow.}
\label{jet-reg}
\end{figure}

\begin{figure}
\epsscale{0.85}
\plotone{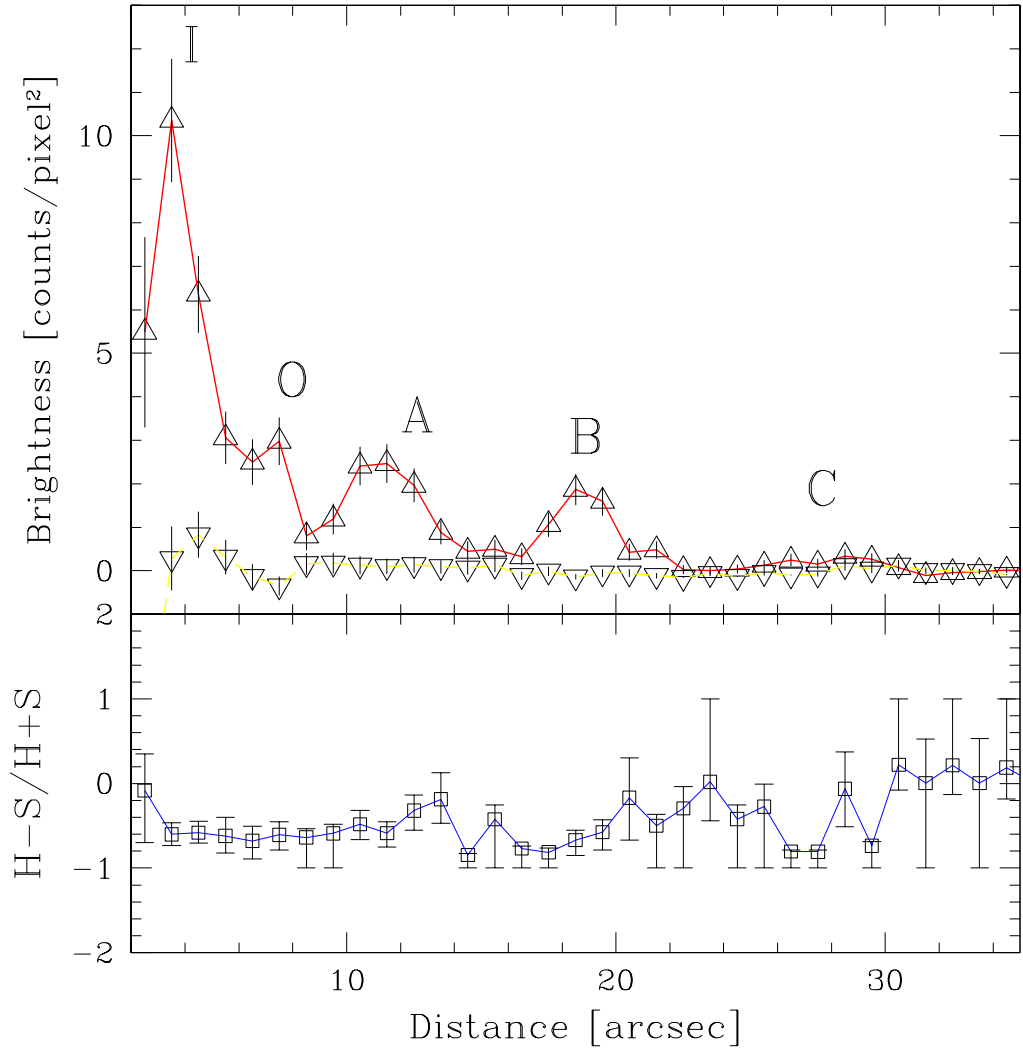}
\caption{Upper panel:  X-ray surface brightness profile of the jet extracted
from the regions shown in Fig.\ref{jet-reg}. The upper triangles mark
the location of the middle point of the box and the profile is
connected with a red solid line.  The control profile is marked with
the down triangles and connected by the dashed yellow line. Photon
energy is within 0.3-7~keV. Units are counts per pixel$^2$ (the
standard ACIS pixel size of 0.492~arcsec). Lower panel: The hardness
ratio along the jet (H-S/H+S) with S are counts between 0.5-2~keV and
H are counts between 2-7~keV. The error bars were calculated using the
BEHR, a Bayesian method described in Park et al (2006).}
\label{profile-hr}
\end{figure}

\begin{figure}
\epsscale{0.85}
\plotone{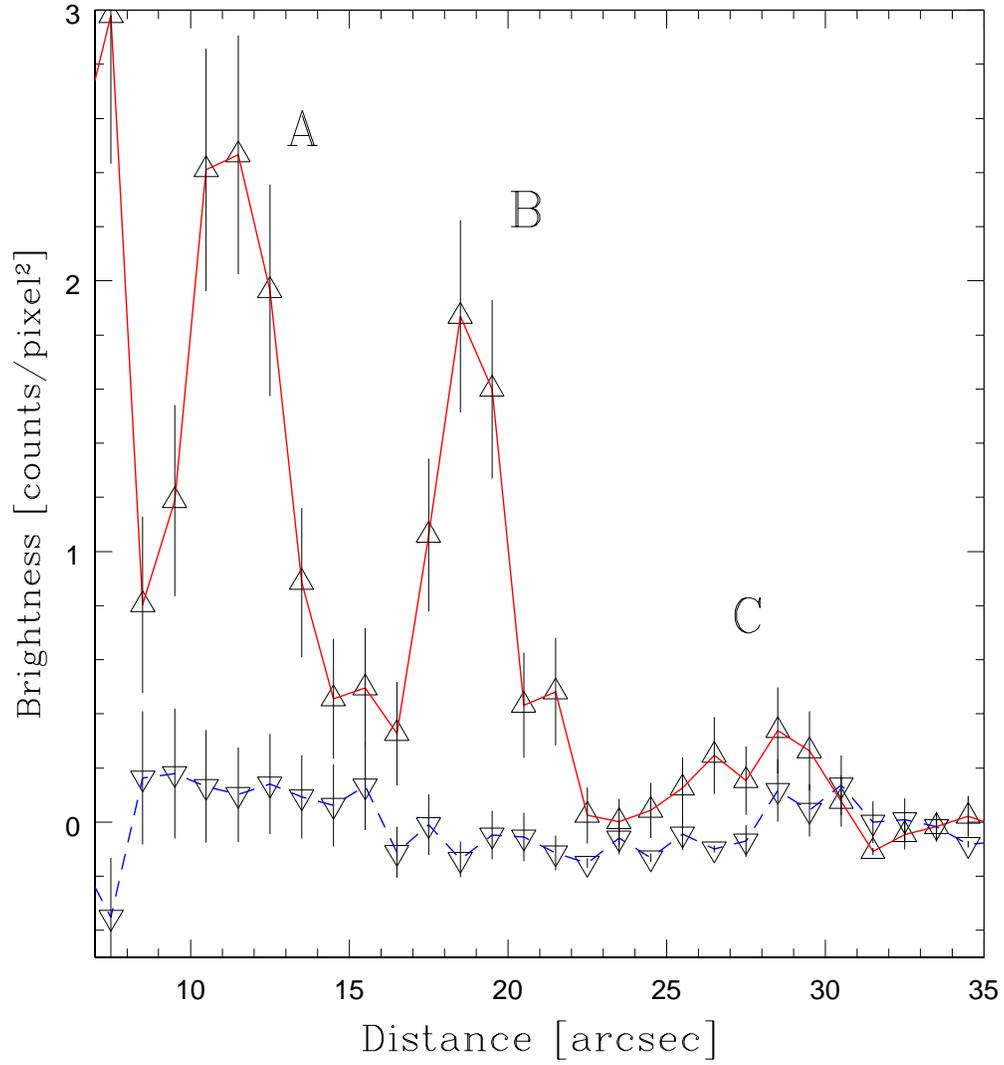}
\caption{The zoom into the three outermost knots of
the jet X-ray surface brightness profile. The jet X-ray surface
brightness profile extracted from the regions shown in
Fig.\ref{jet-reg}.  The triangles mark the location of the middle
point of the box and the profile is connected with a solid red line.
The control profile marked with the down triangles connected by the
dashed blue line.}
\label{profile-zoom}
\end{figure}

\begin{figure}
\epsscale{0.85}
\plotone{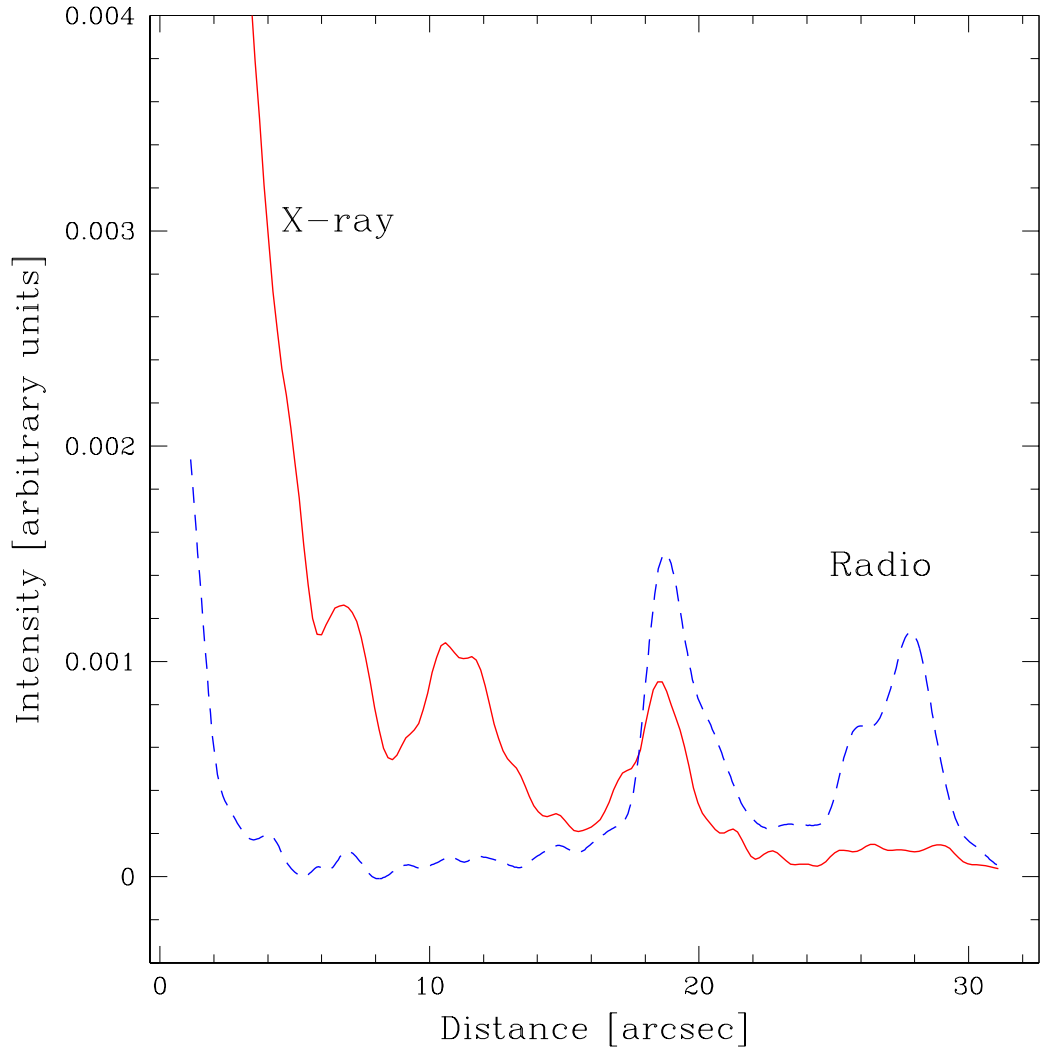}
\caption{Radio (8.5~GHz, dashed line)  and X-ray (0.3-7~keV, solid line) 
brightness along the jet in arbitrary units. 
The X-ray scale has been multiplied by 3.$\times 10^5$ for a visual
comparison of the radio and X-ray profiles. The intensity was
calculated within ds9 using {\tt projection} region of 2.5~arcsec
width along the jet.}
\label{radio-X-profile}
\end{figure}

\begin{figure}
\epsscale{0.85}
\plotone{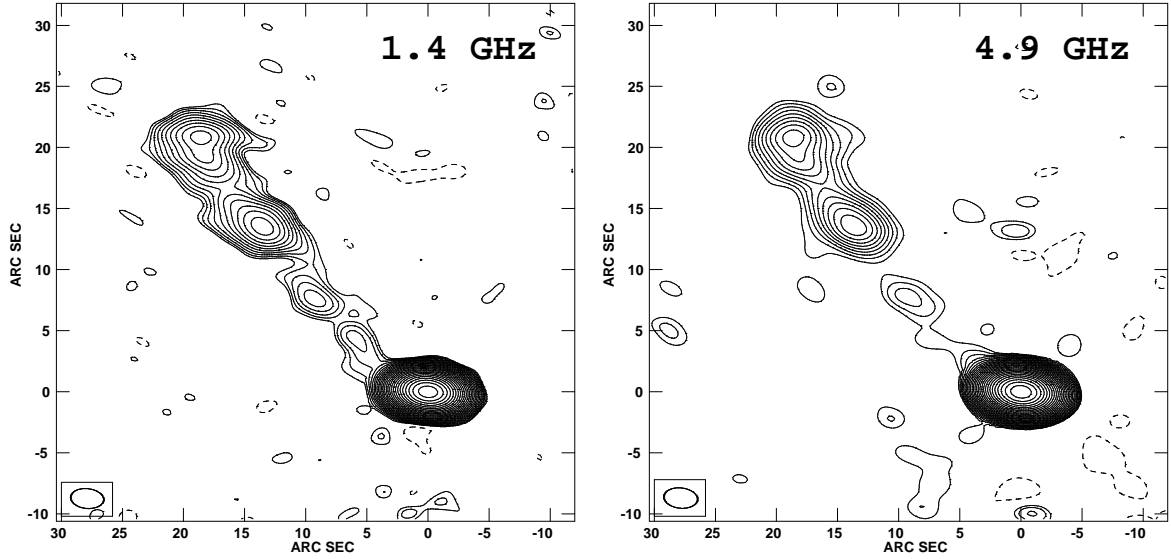}
\caption{VLA 1.4 and 4.9 GHz images of PKS~1127--145. The lowest contour levels
plotted are  0.75 mJy beam$^{-1}$ (1.4 GHz) and
0.39 mJy beam$^{-1}$ (4.9 GHz) and increase by factors of $\sqrt{2}$ up to peaks of
5.46 and 4.14 Jy/beam, respectively. The (uniform weighted) beam sizes are
nearly identical: 2.71\arcsec$\times$1.62\arcsec\ at PA=84.4$^{\circ}$,
and 2.77\arcsec$\times$1.66\arcsec\ at PA=84.0$^{\circ}$, respectively.}
\label{radioCL}
\end{figure}

\begin{figure}
\epsscale{0.85}
\caption{Figure available at {\tt http://hea-www.harvard.edu/$\sim$aneta/PKS1127\_2006/}. A smoothed X-ray image overlayed with the 8.5~GHz radio contour map. 
An X-ray image which has been smoothed with the Gaussian kernel
(1$\sigma$=0.615~arcsec). The radio core has been subtracted. Colors
indicate the X-ray brightness between black and yellow of 10$^{-10}$
and 10$^{-5}$~photons~cm$^{-2}$~s$^{-1}$~pixel$^{-2}$
respectively. The radio contours are plotted between 0.2 and
2.0~mJy/beam, at 0.2, 0.3,0.5, 0.7, 1.0, 1.2,1.3, 1.5,1.7,
2.0~mJy/beam. The radio beam is circular (0.7$''$). The scale is
indicated with the arrows. North is up and East is left.}
\label{radio-X1}
\end{figure}

\begin{figure}
\epsscale{0.85}
\caption{Figure available at {\tt http://hea-www.harvard.edu/$\sim$aneta/PKS112\_2006/}. Outer regions of the jet. A smoothed X-ray image overlayed with 
the 8.5~GHz radio contour map. Colors indicate the X-ray brightness
between black and yellow of 10$^{-10}$ and
10$^{-5}$~photons~cm$^{-2}$~s$^{-1}$~pixel$^{-2}$ respectively. The
radio contours are plotted between 0.2 and 2.0~mJy/beam, at 0.2,
0.3,0.5, 0.7, 1.0, 1.2,1.3, 1.5,1.7, 2.0~mJy/beam.  The radio beam is
circular (0.7$''$). The scale is indicated with the arrows. North is
up and East is left.}
\label{radio-X2}
\end{figure}

\begin{figure}
\epsscale{0.85}
\plotone{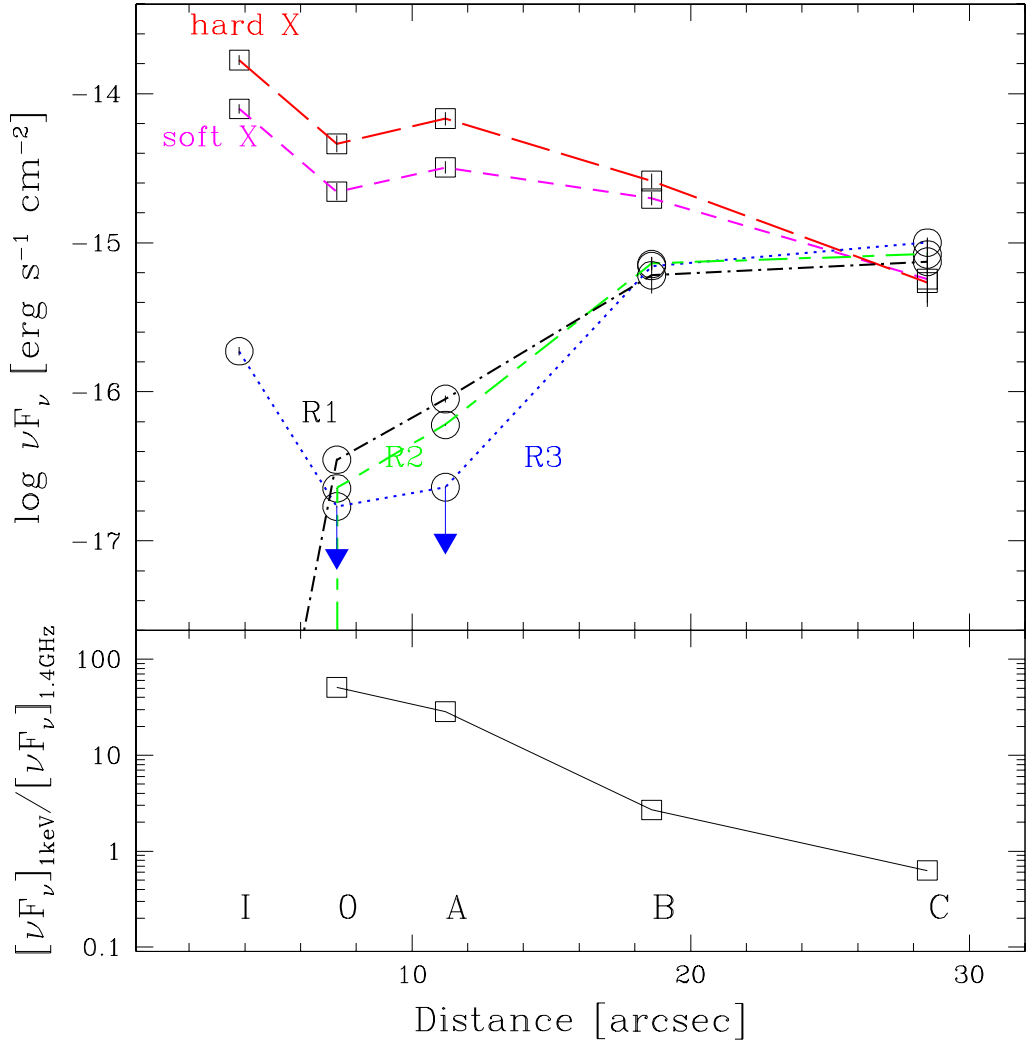}
\caption{Flux along the jet for each observed frequency as indicated on 
the figure. Each curve is plotted with a different style: long dashed
line -- hard X-rays (2-10~keV), short dashed line - soft X-rays (0.5-2
keV), dotted line - R1 = 1.4 GHz, short-long-dashed - R2 = 5~GHz and
dash-dot - R3 = 8.5~GHz. Upper limits are marked with the arrows.
Lower panel shows the flux ratio along the jet.}
\label{profiles}
\end{figure}

\begin{figure}
\epsscale{0.85}
\plotone{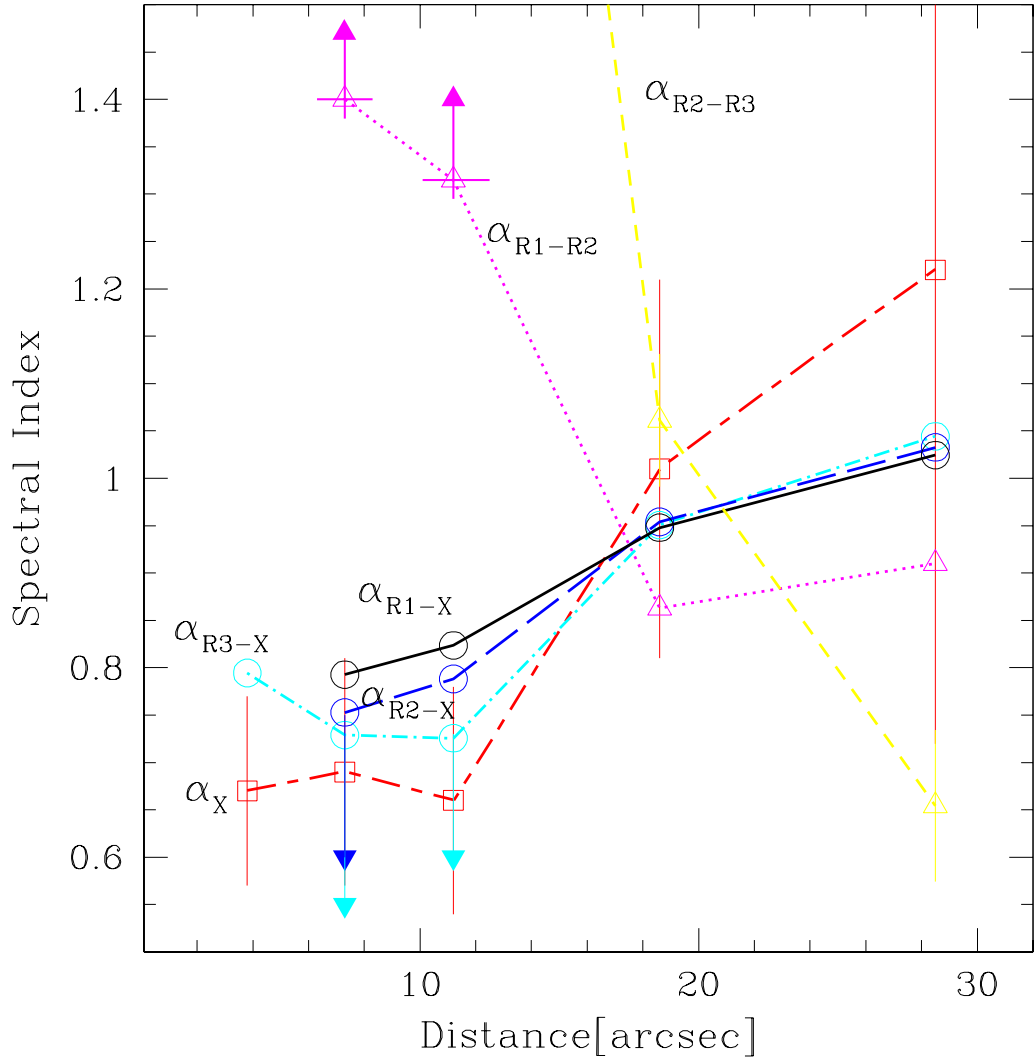}
\caption{Spectral index ($S_{\nu} \sim \nu^{-\alpha}$) 
along the jet for radio and X-ray frequencies. 
Each curve is plotted in a different style and color and it is marked
with symbols: $\alpha_X$- X-ray spectral index, squares, red
long-short dash, $\alpha_{R3-X}$ - index between 8.5~GHz and 1~keV,
circles, cyan dot-dashed line, $\alpha_{R2-X}$ - index between 5~GHz
and 1~keV, circles, blue dashed line, $\alpha_{R1-X}$ - index between
1.4~GHz and 1keV, circles, black solid line, $\alpha_{R1-R2}$ - index
between two radio frequencies 1.4~GHz and 5~GHz, triangles, magenta
dotted line, $\alpha_{R2-R3}$ - index between two radio frequencies
5~GHz and 8.5~GHz, triangles yellow dashed line. Upper and lower 
limits are marked with arrows.}
\label{indices}
\end{figure}

\end{document}